\documentclass[12pt]{article}
	\pdfoutput=1	
\usepackage[utf8]{inputenc}
\usepackage{array} 
\usepackage{amsmath,amssymb}
\usepackage{graphicx}
\usepackage[pdftex]{hyperref}
\usepackage{caption}
\DeclareGraphicsExtensions{.eps,.bmp,.wmf,.jpg}
\numberwithin{equation}{section}
\def\be{\begin{equation}}
\def\ee{\end{equation}}

\def\bea{\begin{eqnarray}}
\def\eea{\end{eqnarray}}

\title{Slow-Roll Inflation with Exponential Potential in Scalar-Tensor Models}
\author{L.N. Granda\thanks{luis.granda@correounivalle.edu.co} ,\, D. F. Jimenez\thanks{jimenez.diego@correounivalle.edu.co}\\{\it Departamento de Fisica, Universidad del Valle}\\{\it A.A. 25360, Cali, Colombia}}
\date{}
\begin{document}
\maketitle

\begin{abstract}
\noindent A study of the slow-roll inflation for an exponential potential in the frame of the scalar-tensor theory is performed, where non-minimal kinetic coupling to curvature and non-minimal coupling of the scalar field to the Gauss-Bonnet invariant are considered. Different models were considered with couplings given by exponential functions of the scalar field, that lead to graceful exit from inflation and give values of the  scalar spectral index and the tensor-to-scalar ratio in the region bounded by the current observational data. Special cases were found, where  the coupling functions are inverse of the potential, that lead to inflation with constant slow-roll parameters, and it was posible to reconstruct the model parameters for given $ns$ and $r$.
In first-order approximation the standard consistency relation maintains its validity in the model with non-minimal coupling, but it modifies in presence of Gauss-Bonnet coupling. The obtained Hubble parameter during inflation, $H\sim 10^{-5} M_p$ and the energy scale of inflation $V^{1/4}\sim 10^{-3} M_p$, are consistent with the upper bounds set by latest observations.

\end{abstract}

\section{\label{intro}Introduction}
The theory of cosmic inflation \cite{guth, linde, steinhardt} that has been favored by the latest observational data \cite{planck13, planck15, planck18, bkp}, is by now the most likely scenario for the early universe,  since it provides the explanation to flatness, horizon and monopole problems, among others, for the standard hot Bing Bang cosmology \cite{revlinde,liddle, riotto, lyth, mukhanov, baumann, nojirioo}. Inflation provides a detailed account of fluctuations that constitute the seeds for the large scale structure and the observed CMB anisotropies \cite{starobinsky1, mukhanov1, starobinsky3, hawking, starobinsky2, guth1, bardeen0, bardeen}, as well as predicts a nearly scale invariant power spectrum.\\
The inflation scenario can be realized by many models, starting from the simplest, the minimally coupled scalar field \cite{linde, steinhardt} and continuing with more elaborated models like non-minimally coupled scalar field \cite{futumase, unruh, barrow1, barrow2, bezrukov}, kinetic inflation \cite{picon}, vector inflation \cite{ford, koivisto, mukhanov2}, inflaton potential in supergravity \cite{kawasaki, davies, kallosh}, string theory inspired inflation \cite{soda1, soda2, maldacena, kallosh1, soda3, baumann2}, Dirac-Born-Infeld inflation model \cite{silver, silver1, chen, easson}, $\alpha$-attractor models originated in supergravity \cite{linde3, ferrara, sergei1, dimopoulos, akrami}. Apart from the DBI models of inflation, another class of ghost-free models has been recently considered, named "Galileon" models \cite{nicolis, deffayet}. 
The main characteristic of these models is that the gravitational and scalar field equations remain as second-order differential equations. 
The Galileon terms modify the kinetic term compared to the standard canonical scalar field, which in turn can relax the physical constraints on the potential. In the case of the Higgs-type potential, for instance, one of the effects of the higher derivative terms is the reduction of the self coupling of the Higgs boson, so that the spectra of primordial density perturbations are consistent with the present observational data \cite{kamada, ohashi}. 
Galilean models of inflation have been considered in \cite{kamada, ohashi, yokoyama, mizuno, burrage, kobayashi1}. Some aspects of slow-roll inflation with non-minimal kinetic coupling have been analyzed in \cite{capozziello1, germani, granda3, stsujikawa, nanyang, gansu}. 
For a sample papers devoted to the study of slow-roll inflation in the context of Gauss-Bonnet (GB) coupling see  \cite{Kanti:1998jd, soda4, Guo:2009uk, Guo:2010jr, Jiang:2013gza, Koh:2014bka, Kanti:2015pda, soda5, vandeBruck:2017voa, quiang, sergeioik, sergeioik1, chakraborty, tanmoy}.\\ 
\noindent This paper is dedicated to the study of the slow-roll inflation in the scalar-tensor model with non-minimal kinetic coupling to the Einstein tensor and coupling of the scalar field to the Gauss-Bonnet 4-dimensional invariant. The non-minimal couplings of the scalar field to curvature of the type considered in the present paper arise, among other couplings, in fundamental theories like supergravity and string theory after specific compactification to an effective four dimensional theory \cite{fradkin, gross, tseytlin, meissner, cartier}, where the scalar field is related to the size of the compact extra dimensions and the (exponential) potential is related to the curvature of the extra dimensional manifold. 
Note however that for the potential  $V=V_0 e^{\lambda\phi}$, that  produces power-law $a\propto t^{2/\lambda^2}$, the constant $\lambda$ that appear from compactifications is usually of order $1$ or greater, which is insufficient to generate inflation. By considering non-minimal derivative and Gauss-Bonnet couplings, this problem can be avoided. This makes it appealing  to analyze the mechanism of slow-roll inflation in such  theories where the scalar field appears non-minimally coupled to curvature terms. This could provide a connection with fundamental  theories in a high curvature regime characteristic of inflation.\\
The  Gauss-Bonnet and non-minimal kinetic couplings have also been considered in the dark energy problem  \cite{nojiriodintsov, neupane, koivisto1, koivisto2, suchkov, lgranda}. 
Particularly, couplings  with exponential form have been considered in \cite{grandaloaiza} to study the late time cosmological dynamics, where stable and saddle scaling solutions have been obtained and a critical points corresponding to de Sitter solution were found.  The de Sitter solutions correspond to the critical points {\bf C} and {\bf E} in \cite{grandaloaiza} with marginal stability, which could be considered as possible (saddle point)  inflationary solutions. So this model could describe an inflationary de Sitter solution that can evolve towards scaling solutions with saddle character or to stable attractor dominated by the scalar field, describing accelerated expansion \cite{grandaloaiza}. Thus, the present model can provide a connection between early time inflation and late time accelerated expansion (in \cite{granjimsan} special cases of quintessence and phantom solutions have been studied with exponential couplings). For unified description of early time inflation and late time accelerated expansion in the framework of scalar tensor theories see \cite{sergeinojiri}. \\
In the present work we consider models with exponential potential and exponential couplings. An important feature of the exponential potential (in the framework of minimally coupled scalar field model) is that under its dominance the universe expands following a power-law, which describes the asymptotic behavior of the background spacetime in different epochs. This is the case of the late time dark energy dominated universe, where the exponential potential can give rise to accelerated expansion  \cite{copeland1, copeland2}.
Applied to the study of the early universe, the exponential potential in the minimally coupled scalar field model, gives rise to power-law inflation \cite{abbott, lucchin, yokoyama1, liddle1, ratra} with constant slow-roll parameters. This implies that the exponential potential lacks a successful exit  from inflation, which added to the fact that the tensor-to-scalar ratio is larger than the limits set by Planck data, rules out the exponential potential in the standard canonical scalar field model. In the present paper we address the above shortcomings of the exponential potential, this time in the frame of scalar-tensor theories, taking into account non-minimal kinetic an GB couplings, which could  play relevant role in the high curvature regime typical for inflation. We find that the above couplings predict values for the scalar spectral index and the tensor-to-scalar ratio that fall in the region quoted by the latest observational data.
The paper is organized as follows. In the next section we introduce the model, the background field equations and define the slow-roll parameters. In section 3 we use quadratic action for the scalar and tensor perturbations to evaluate the primordial power spectra. In section 4 we analyze several models with exponential potential and exponential couplings. Some discussion is presented in section 5.
\section{The model and background equations}
We consider the following scalar-tensor model
\be\label{eqm1}
S=\int d^4x\sqrt{-g}\left[\frac{1}{2}F(\phi)R-\frac{1}{2}\partial_{\mu}\phi\partial^{\mu}\phi-V(\phi)+F_1(\phi)G_{\mu\nu}\partial^{\mu}\phi\partial^{\nu}\phi-F_2(\phi){\cal G}\right]
\ee
where $G_{\mu\nu}$ is the Einstein's tensor, ${\cal G}$ is the GB 4-dimensional invariant given by
\be\label{eqm2}
{\cal G}=R^2-4R_{\mu\nu}R^{\mu\nu}+R_{\mu\nu\lambda\rho}R^{\mu\nu\lambda\rho}
\ee
\be\label{eqm3}
F(\phi)=\frac{1}{\kappa^2}+f(\phi),
\ee
and $\kappa^2=M_p^{-2}=8\pi G$.
One remarkable characteristic of this model is that it yields second-order field equations and can avoid Ostrogradski instabilities. In the spatially flat FRW background 
\be\label{eqm4}
ds^2=-dt^2+a(t)^2\left(dx^2+dy^2+dz^2\right),
\ee
one can write the field equations as follows
\be\label{eqm5}
3H^2F\left(1-\frac{3F_1\dot{\phi}^2}{F}-\frac{8H\dot{F_2}}{F}\right)=\frac{1}{2}\dot{\phi}^2+V-3H\dot{F}
\ee
\be\label{eqm6}
\begin{aligned}
2\dot{H}F\left(1-\frac{F_1\dot{\phi}^2}{F}-\frac{8H\dot{F_2}}{F}\right)&=-\dot{\phi}^2-\ddot{F}+H\dot{F}+8H^2\ddot{F_2}-8H^3\dot{F_2}\\
&-6H^2F_1\dot{\phi}^2+4HF_1\dot{\phi}\ddot{\phi}+2H\dot{F}_1\dot{\phi}^2
\end{aligned}
\ee
\be\label{eqm7}
\begin{aligned}
\ddot{\phi}+3H\dot{\phi}&+V'-3F'\left(2H^2+\dot{H}\right)+24H^2\left(H^2+\dot{H}\right)F_2'+18H^3F_1\dot{\phi}\\&+12H\dot{H}F_1\dot{\phi}+6H^2F_1\ddot{\phi}+3H^2F_1'\dot{\phi}^2=0
\end{aligned}
\ee
where ($'$) denotes derivative with respect to the scalar field. Related to the different terms in the action (\ref{eqm1}) we define the following slow-roll parameters 
\be\label{eqm8}
\epsilon_0=-\frac{\dot{H}}{H^2},\;\;\; \epsilon_1=\frac{\dot{\epsilon}_0}{H\epsilon_0}
\ee
\be\label{eqm9}
\ell_0=\frac{\dot{F}}{HF},\;\;\; \ell_1=\frac{\dot{\ell}_0}{H\ell_0}
\ee
\be\label{eqm10}
k_0=\frac{3F_1\dot{\phi}^2}{F},\;\;\; k_1=\frac{\dot{k}_0}{Hk_0}
\ee
\be\label{eqm11}
\Delta_0=\frac{8H\dot{F_2}}{F},\;\;\; \Delta_1=\frac{\dot{\Delta}_0}{H\Delta_0}
\ee
The slow-roll conditions in this model are satisfied if $\epsilon_0,\; \epsilon_1,\; \Delta_0,....<<1$.
From the cosmological equations (\ref{eqm5}) and (\ref{eqm6}) and using the parameters (\ref{eqm8})-(\ref{eqm11}) we can write the following expressions for $\dot{\phi}^2$ and $V$
\be\label{eqm12}
\begin{aligned}
V=&H^2F\Big[3-\frac{5}{2}\Delta_0-2k_0-\epsilon_0+\frac{5}{2}\ell_0+\frac{1}{2}\ell_0\left(\ell_1-\epsilon_0+\ell_0\right)\\&-\frac{1}{2}\Delta_0\left(\Delta_1-\epsilon_0+\ell_0\right)-\frac{1}{3}k_0\left(k_1+\ell_0-\epsilon_0\right)\Big]
\end{aligned}
\ee
\be\label{eqm13}
\begin{aligned}
\dot{\phi}^2=&H^2F\Big[2\epsilon_0+\ell_0-\Delta_0-2k_0+\Delta_0\left(\Delta_1-\epsilon_0+\ell_0\right)-\\& \ell_0\left(\ell_1-\epsilon_0+\ell_0\right)+\frac{2}{3}k_0\left(k_1+\ell_0-\epsilon_0\right)\Big]
\end{aligned}
\ee
where we used
\be\label{eqm13a}
\ddot{F}=H^2F\ell_0\left(\ell_1-\epsilon_0+\ell_0\right),\;\;\; \ddot{F}_2=\frac{F\Delta_0}{8}\left(\Delta_1+\epsilon_0+\ell_0\right)
\ee
It is also useful to define the variable $Y$ from Eq. (\ref{eqm13}) as
\be\label{eqm13b}
Y=\frac{\dot{\phi}^2}{H^2F}
\ee
where it follows that $Y = {\cal O}(\varepsilon )$. 
Under the slow-roll conditions $\ddot{\phi}<<3H\dot{\phi}$ and $\ell_i, k_i, \Delta_i<<1$, it follows from the field equations (\ref{eqm5})-(\ref{eqm7}) that they can be reduced to
\be\label{eqm14}
3H^2F\simeq V,
\ee
\be\label{eqm15}
2\dot{H}F\simeq -\dot{\phi}^2+H\dot{F}-6H^2F_1\dot{\phi}^2-8H^3\dot{F}_2,
\ee
\be\label{eqm16}
3H\dot{\phi}+V'-6H^2F'+18H^3F_1\dot{\phi}+24H^4F'_2\simeq 0,
\ee
The scalar field equation (\ref{eqm16}) allows to determine the number of $e$-folds as
\be\label{eqm17}
N=\int_{\phi_I}^{\phi_E}\frac{H}{\dot{\phi}}d\phi=\int_{\phi_I}^{\phi_E}\frac{H^2+6H^4F_1}{2H^2F'-8H^4F'_2-\frac{1}{3}V'}d\phi
\ee
where $\phi_I$ and $\phi_E$ are the values of the scalar field at the beginning and end of inflation respectively. 
\section{Second order action for the scalar and tensor perturbations}
{\bf Scalar Perturbations}.\\

\noindent The details of the first and second order perturbations fro the model (\ref{eqm1}) are given in \cite{grajim}. The second order action for the scalar perturbations is given by the following expression
\be\label{slr1}
\delta S_s^{2}=\int dt d^3xa^3\left[{\cal G}_s\dot{\xi}^2-\frac{{\cal F}_s}{a^2}\left(\nabla\xi\right)^2\right]
\ee
where
\be\label{slr2}
{\cal G}_s=\frac{\Sigma}{\Theta^2}{\cal G}_T^2+3{\cal G}_T
\ee
\be\label{slr3}
{\cal F}_s=\frac{1}{a}\frac{d}{dt}\left(\frac{a}{\Theta}{\cal G}_T^2\right)-{\cal F}_T
\ee
with
\be\label{slr4}
{\cal G}_T=F-F_1\dot{\phi}^2-8H\dot{F}_2.
\ee
\be\label{slr5}
{\cal F}_T=F+F_1\dot{\phi}^2-8\ddot{F}_2
\ee
\be\label{slr6}
\Theta=FH+\frac{1}{2}\dot{F}-3HF_1\dot{\phi}^2-12H^2\dot{F}_2
\ee
\be\label{slr7}
\Sigma=-3FH^2-3H\dot{F}+\frac{1}{2}\dot{\phi}^2+18H^2F_1\dot{\phi}^2+48H^3\dot {F}_2
\ee
And the sound speed of scalar perturbations is given by 
\be\label{slr8}
c_S^2=\frac{{\cal F}_S}{{\cal G}_S}
\ee
The conditions for avoidance of ghost and Laplacian instabilities as seen from the action (\ref{slr1}) are
$$ {\cal F}>0,\;\;\;\; {\cal G}>0 $$
We can rewrite ${\cal G}_T$, ${\cal F}_T$, $\Theta$ and $\Sigma$ in terms of the slow-roll parameters (\ref{eqm8})-(\ref{eqm11}) and using Eqs. (\ref{eqm13}) and (\ref{eqm13a}), as follows
\be\label{slr8a}
{\cal G}_T=F\left(1-\frac{1}{3}k_0-\Delta_0\right)
\ee
\be\label{slr8b}
{\cal F}_T=F\left(1+\frac{1}{3}k_0-\Delta_0\left(\Delta_1+\epsilon_0+\ell_0\right)\right)
\ee
\be\label{slr8c}
\Theta=FH\left(1+\frac{1}{2}\ell_0-k_0-\frac{3}{2}\Delta_0\right)
\ee
\be\label{slr8d}
\begin{aligned}
\Sigma=&-FH^2\Big[3-\epsilon_0+\frac{5}{2}\ell_0-5k_0-\frac{11}{2}\Delta_0+\frac{1}{2}\ell_0\left(\ell_1-\epsilon_0+\ell_0\right)\\&-\frac{1}{3}k_0\left(k_1-\epsilon_0+\ell_0\right)-\frac{1}{2}\Delta_0\left(\Delta_1-\epsilon_0+\ell_0\right)\Big]
\end{aligned}
\ee
The expressions for ${\cal G}_S$ and $c_S^2$ in terms of the slow roll parameters can be written as
\be\label{slowGS}
{{\cal G}_S} = \frac{{F\left( {\frac{1}{2}Y + {k_0} + \frac{3}{4}{W^2}(1 - {\Delta _0} - \frac{1}{3}{k_0})} \right)}}{{{{\left( {1 + \frac{1}{2}W} \right)}^2}}}
\ee
\be\label{slowCS}
c_S^2 = 1 + \frac{{{W^2}\left( {\frac{1}{2}{\Delta _0}({\Delta _1} + {\varepsilon _0} + {l_0} - 1) - \frac{1}{3}{k_0}} \right) + W\left( {\frac{2}{3}{k_0}\left( {2 - {k_1} - {l_0}} \right) + 2{\Delta _0}{\varepsilon _0}} \right) - \frac{4}{3}{k_0}{\varepsilon _0}}}{{Y + 2{k_0} + \frac{3}{2}{W^2}(1 - {\Delta _0} - \frac{1}{3}{k_0})}}
\ee
where

\be\label{slowW}
W = \frac{\ell_0-\Delta_0-\frac{4}{3}k_0}{1-\Delta_0-\frac{1}{3}k_0}
\ee
Notice that in general ${\cal G}_S=F {\cal O}(\varepsilon )$ and $c_S^2 = 1 +{\cal O}(\varepsilon )$. Also in absence of the kinetic coupling it follows that $c_S^2 = 1 +{\cal O}(\varepsilon^2 )$. Keeping first order terms in slow-roll parameters, the expressions for ${\cal G}_S$ y $c_S^2$ reduce to 
\be\label{aproxGS}
{\cal G}_S = F\left( {{\varepsilon _0} + \frac{1}{2}{l_0} - \frac{1}{2}{\Delta _0}} \right)
\ee
\be\label{aproxCS}
c_S^2 = 1 + \frac{{\frac{4}{3}{k_0}\left( {{l_0} - {\Delta _0} - \frac{4}{3}{k_0}} \right) - \frac{4}{3}{k_0}{\varepsilon _0}}}{{2{\varepsilon _0} + {l_0} - {\Delta _0}}}
\ee 
\noindent After the appropriate change of variables  to normalize the action (\ref{eqm1}), we find the equation of motion, working in the Fourier representation, as (\cite{kobayashi1}) (see (F.7) of \cite{grajim})
\be\label{slr12}
\tilde{U}_{\vec{k}}''+\left(k^2-\frac{\tilde{z''}}{\tilde{z}}\right)\tilde{U}_{\vec{k}}=0
\ee
where 
\be\label{slr9}
d\tau_s=\frac{c_S}{a}dt,\;\;\, \tilde{z}=\sqrt{2}a\left({\cal F}_S{\cal G}_S\right)^{1/4},\;\;\; \tilde{U}=\xi\tilde{z}
\ee
From (\ref{slr9}), and keeping up to first-order terms in slow-roll variables in (\ref{slowGS}) and (\ref{slowCS}), we find the following expression for $\tilde{z}''/\tilde{z}$ 
\be\label{slr14}
\frac{\tilde{z}''}{\tilde{z}}=\frac{a^2H^2}{c_S^2}\Big[2-\epsilon_0+\frac{3}{2}\ell_0+\frac{3}{2}\frac{2\epsilon_0\epsilon_1+\ell_0\ell_1-\Delta_0\Delta_1}{2\epsilon_0+\ell_0-\Delta_0}\Big].
\ee
Taking into account the slow-roll parameters we can rewrite the Eq. (\ref{slr12}) in the form
\be\label{slr20}
\tilde{U}_k''+k^2\tilde{U}_{k}+\frac{1}{\tau_s^2}\left(\mu_s^2-\frac{1}{4}\right)\tilde{U}_{k}=0
\ee
where
\be\label{slr21}
\mu_s^2=\frac{9}{4}\left[1+\frac{4}{3}\epsilon_0+\frac{2}{3}\ell_0+\frac{2}{3}\frac{2\epsilon_0\epsilon_1+\ell_0\ell_1-\Delta_0\Delta_1}{2\epsilon_0+\ell_0-\Delta_0} \right],
\ee
After the integration of (\ref{slr20}) using the slow-roll formalism (see \cite{grajim} for details) we find, at super horizon scales ($c_S k<<aH$), the following asymptotic solution 
\be\label{slr24}
\tilde{U}_k=\frac{1}{\sqrt{2}}e^{i\frac{\pi}{2}(\mu_s-\frac{1}{2})}2^{\mu_s-\frac{3}{2}}\frac{\Gamma(\mu_s)}{\Gamma(3/2)}\sqrt{-\tau_s}(-k\tau_s)^{-\mu_s}.
\ee
On the other hand, from the relationship
\be\label{slr25}
\frac{\tilde{z}'}{\tilde{z}}=-\frac{1}{(1-\epsilon_0)\tau_s}\Big[1+\frac{1}{2}\ell_0+\frac{1}{2}\frac{2\epsilon_0\epsilon_1+\ell_0\ell_1-\Delta_0\Delta_1}{2\epsilon_0+\ell_0-\Delta_0}\Big]=-\frac{1}{\tau_s}\left(\mu_s-\frac{1}{2}\right),
\ee
and after integrating in the slow-roll approximation we find
\be\label{slr26}
\tilde{z}\propto \tau_s^{\frac{1}{2}-\mu_s},
\ee
which gives in the super horizon regime, from from (\ref{slr9}), the following $k$-dependence for the amplitude of the scalar perturbations  
\be\label{slr27}
\xi_k=\frac{\tilde{U}_k}{\tilde{z}}\propto k^{-\mu_s}
\ee
Then, from the power spectra for the scalar perturbations 
\be\label{slr28}
P_{\xi}=\frac{k^3}{2\pi^2}|\xi_k|^2
\ee
we find the spectral index, in first order in slow-roll parameters
\be\label{slr29}
n_s-1=\frac{d\ln P_{\xi}}{d\ln k}=3-2\mu_s=-2\epsilon_0-\ell_0-\frac{2\epsilon_0\epsilon_1+\ell_0\ell_1-\Delta_0\Delta_1}{2\epsilon_0+\ell_0-\Delta_0}
\ee

\noindent {\bf Tensor perturbations}.\\

\noindent The second order action for the tensor perturbations is given by (\cite{grajim})
\be\label{slrt1}
\delta S_2=\frac{1}{8}\int d^3xdt{\cal G}_T a^2\left[\left(\dot{h}_{ij}\right)^2-\frac{c_T^2}{a^2}\left(\nabla h_{ij}\right)^2\right]
\ee 
where ${\cal G}_T$ and ${\cal F}_T$ are defined in (\ref{slr4}) and (\ref{slr5}) (in terms of the slow-roll variables (\ref{eqm8})-(\ref{eqm11})). The velocity of tensor perturbations is given by
\be\label{slrt2}
c_T^2=\frac{{\cal F}_T}{{\cal G}_T}=\frac{3+k_0-3\Delta_0\left(\Delta_1+\epsilon_0+\ell_0\right)}{3-k_0-3\Delta_0}.
\ee
Following the same lines as for the scalar perturbations and introducing the following variables
\be\label{slrt3}
d\tau_T=\frac{c_T}{a}dt,\;\;\; z_T=\frac{a}{2}\left({\cal F}_T{\cal G}_T\right)^{1/4},\;\;\; v_{ij}=z_T h_{ij},
\ee
that lead to the equation
\be\label{slrt6}
v''_{(k)ij}+\left(k^2-\frac{z''_T}{z_T}\right)v_{(k)ij}=0,
\ee
The deduction of the power spectrum for primordial tensor perturbations follows the same pattern as for the scalar perturbations. At super horizon scales ($c_Tk<<aH$) the tensor modes (\ref{slrt1}) have the same functional form for the asymptotic behavior as the scalar modes (\ref{slr24}), and therefore we can write power spectrum for tensor perturbations as
\be\label{slrt11}
P_T=\frac{k^3}{2\pi^2}|h^{(k)}_{ij}|^2
\ee
where, in first order in slow-roll parameters, the tensor spectral index has the following form \cite{grajim}
\be\label{slrt11}
n_T=3-2\mu_T=-2\epsilon_0-\ell_0
\ee
where
\be\label{slr11a}
\mu_T=\frac{3}{2}+\epsilon_0+\frac{1}{2}\ell_0.
\ee
An important quantity is the relative contribution to the power spectra of tensor and scalar perturbations, defined as the tensor/scalar ratio $r$
\be\label{slrt12}
r=\frac{P_T(k)}{P_{\xi}(k)}.
\ee
For the scalar perturbations, using (\ref{slr28}), we can write the power spectra as
\be\label{slrt13}
P_{\xi}=A_S\frac{H^2}{(2\pi)^2}\frac{{\cal G}_S^{1/2}}{{\cal F}_S^{3/2}}
\ee
where 
$$A_S=\frac{1}{2}2^{2\mu_s-3}\Big|\frac{\Gamma(\mu_s)}{\Gamma(3/2)}\Big|^2$$
and all magnitudes are evaluated at the moment of horizon exit when $c_s k=aH$ ($k\tau_s=-1$). For $\tilde{z}$ we used (\ref{slr9}) with $a=c_S k/H$. In analogous way we can write the power spectra for tensor perturbations as
\be\label{slrt14}
P_T= 16A_T\frac{H^2}{(2\pi)^2}\frac{{\cal G}_T^{1/2}}{{\cal F}_T^{3/2}}
\ee
where 
$$A_T=\frac{1}{2}2^{2\mu_T-3}\Big|\frac{\Gamma(\mu_T)}{\Gamma(3/2)}\Big|^2.$$
Noticing that $A_T/A_S\simeq 1$ when evaluated at the limit $\epsilon_0,\ell_0,\Delta_0,...<<1$, as follows from (\ref{slr21}) and (\ref{slr11a}), we can write the tensor/scalar ratio as follows
\be\label{slrt15}
r=16\frac{{\cal G}_T^{1/2}{\cal F}_S^{3/2}}{{\cal G}_S^{1/2}{\cal F}_T^{3/2}}=16\frac{c_S^3{\cal G}_S}{c_T^3{\cal G}_T}
\ee
taking into account the expressions for ${\cal G}_T, {\cal F}_T, {\cal G}_S,{\cal F}_S$ given in (\ref{slr8a})-(\ref{slowW}), up to first order, and using the condition $\epsilon_0,\ell_0,k_0,\Delta_0<<1$, then  we can see that $c_T\simeq c_S\simeq 1$ (in fact in the limit $\ell_0\rightarrow 0$, $c_S=1$ independently of the values of $\epsilon_0$ and $\Delta_0$) and we can make the approximation
\be\label{slrt16}
r=8\left(\frac{2\epsilon_0+\ell_0-\Delta_0}{1-\frac{1}{3}k_0-\Delta_0}\right)\simeq 8\left(2\epsilon_0+\ell_0-\Delta_0\right)
\ee
which is a modified consistency relation due to the non-minimal and GB couplings. In the limit $\ell_0,\Delta_0\rightarrow 0$ it gives the standard consistency relation for the single canonical scalar field inflation 
\be\label{slrt17}
r= -8n_T,
\ee
with $n_T=-2\epsilon_0$. Note that if the model contains only non-minimal coupling $F(\phi)$, then $r\simeq 8(2\epsilon_0+\ell_0)$, and from (\ref{slrt11}) it follows that the standard consistency relation (\ref{slrt17}) remains valid in presence of non-minimal coupling. 
In the general case from (\ref{slrt16}) we find the deviation from the standard consistency relation in the form
\be\label{slrt18}
r=-8n_T+\delta r,\;\;\; \delta r=-8\Delta_0,
\ee
with $n_T$ given by (\ref{slrt11}). Here for standard consistency relation we mean the relation (\ref{slrt17}) independently of the content of $n_T$. This expression can also be written as
\be\label{consistency} 
r=-8n_T\left(1+\frac{\Delta_0}{n_T}\right)=-8n_T\left(1-\frac{\Delta_0}{2\epsilon_0+\ell_0}\right)=-8\gamma n_T
\ee
where $\gamma=1-\Delta_0/(2\epsilon_0+\ell_0)$  characterizes the deviation from the standard consistency relation. Note that this deviation in first-order approximation is independent of $k_0$.
Thus, the consistency relation still valid in the case of non-minimal coupling ($\Delta_0=0$), and a deviation from the standard consistency relation can reveal the effect of interactions beyond the simple canonical or non-minimally coupled scalar field. 


\section{Inflation Driven by Exponential Potential and Exponential Couplings}
The exponential potential leads to scaling solutions important to describe different epochs of cosmological evolution, including solutions with accelerated expansion. In the standard minimally coupled scalar field it leads to inflationary solutions with constant slow-roll parameters, that lead to eternal inflation, which added to the strong signal of gravitational waves ($r>0.1$), makes the model inviable. As stated in the introduction, the exponential potential and couplings appear in a number of compactifications from higher dimensional fundamental theories  such as supergravity and string theory, where the scalar field encodes the size of the extra dimensions.  Although these couplings are inspired by higher-dimensional gravitational theories, in the present study we are not trying to  match any specific model coming directly from higher dimensional compactifications. The viability of the present model is probed by the fact that it leads to graceful exit from inflation, and after estimating the power spectra of scalar and tensor perturbations it gives the main inflationary observables $n_s$ and $r$ in the region quoted by the latest observational data. \\

\noindent {\bf Kinetic coupling.}\\

\noindent Let us start with the model (\ref{eqm1}) with the explicit form of the couplings given by\\
\be\label{expo1}
F\left( \phi  \right) = \frac{1}{{{\kappa ^2}}},\,\,\,\,\,\,\,\,\,V\left( \phi  \right) = V_0 e^{-\lambda\kappa\phi},\,\,\,\,\,\,\,\,\,{F_1}\left( \phi  \right) =f_{k} e^{-\eta\kappa\phi} ,\,\,\,\,\,\,\,\,{F_2}\left( \phi  \right) = 0.
\ee
from (\ref{eqm8})-(\ref{eqm11}), using (\ref{eqm14})-(\ref{eqm16}) we find the slow-roll parameters
\be\nonumber
\epsilon_0=\frac{\lambda^2}{2\left(2\alpha e^{-(\lambda+\eta)\phi}+1\right)},\;\;\; \epsilon_1=\frac{2\alpha\lambda(\lambda+\eta)e^{(\lambda+\eta)\phi}}{\left(e^{(\lambda+\eta)\phi}+2\alpha\right)^2},
\ee
\be\label{epsilons}
k_0=\frac{\alpha\lambda^2 e^{(\lambda+\eta)\phi}}{\left(e^{(\lambda+\eta)\phi}+2\alpha\right)^2},\;\;\; k_1=-\frac{\lambda(\lambda+\eta)e^{(\lambda+\eta)\phi}\left(e^{(\lambda+\eta)\phi}-2\alpha\right)}{\left(e^{(\lambda+\eta)\phi}+2\alpha\right)^2}
\ee
where we have set $\kappa=1$ and $\alpha=V_0f_k$. In standard slow-roll inflation ($f_k=0,\; \eta=0$) the condition $\lambda^2<<1$ is required, while in the presence of kinetic coupling this condition can be avoided due to the $\phi$ dependence in the slow-roll parameters. This $\phi$-dependence of the slow-roll parameters also allows the graceful exit from inflation. 
Using the condition $\epsilon_0(\phi_E)=1$ we find the expression for the scalar field at the end of inflation as
\be\label{fielde}
\phi_E=\frac{1}{\lambda+\eta}\ln\left[\frac{4\alpha}{\lambda^2-2}\right].
\ee 
With $f_k$ being positive, this field is well defined whenever $\lambda>\sqrt{2}$. It is clear from this expression that the larger $\eta$, the smaller $\phi_E$ can be. It also follows that  $\phi_E$ varies very slowly with the increment of $\alpha$ because of the logarithm dependence. Assuming for instance $\lambda=2, \eta=5, \alpha=10^3$, give $\phi_E\simeq 1.08 M_p$, and $\lambda=2, \eta=5, \alpha=10^2$ give $\phi_E\simeq 0.76 M_p$.\\
The Eq. (\ref{eqm17}) gives the number of e-foldings as
\be\label{efolds}
N=\frac{1}{2\lambda(\lambda+\eta)}\left(2-\lambda^2+2\ln\left[\frac{4\alpha}{\lambda^2-2}\right]\right)-\left(\frac{\phi_I}{\lambda}-\frac{2\alpha e^{-(\lambda+\eta)\phi_I}}{\lambda(\lambda+\eta)}\right)
\ee
where $\phi_I$ is the scalar field $N$ $e$-folds before the end of inflation. Solving this equation gives the explicit form of $\phi_I$
\be\label{fieldi}
\phi_I=\frac{1}{2(\lambda+\eta)}\left[2\ln\left(\frac{4\alpha}{\lambda^2-2}\right)-2\lambda N\left(\lambda+\eta\right)-\lambda^2+2W\left[\frac{1}{2}\left(\lambda^2-2\right)e^{\frac{\lambda^2}{2}+\lambda N(\lambda+\eta)-1}\right]\right].
\ee
For the scalar spectral index, we see from (\ref{slr29}) that up to first order in slow-roll parameters $n_s$ does not depend on $k_0$ and $k_1$. So, if the model contains only non-minimal kinetic coupling the scalar spectral index becomes $n_s=1-2\epsilon_0-\epsilon_1$, and for the same reason from (\ref{slrt16}) follows that $r=16\epsilon_0$. However, both $\epsilon_0$ and $\epsilon_1$ depend on all the parameters of the model. The analytical expression for $n_s$ is given by 
\be\label{ns1}
n_s=1-\frac{\lambda^2}{2\alpha e^{-(\lambda+\eta)\phi_I}+1}-\frac{2\alpha\lambda(\lambda+\eta)e^{(\lambda+\eta)\phi_I}}{\left(e^{(\lambda+\eta)\phi_I}+2\alpha\right)^2},
\ee
where $\phi_I$ is given by (\ref{fieldi}). And for the tensor-to-scalar-ratio it is found
\be\label{r1}
r=\frac{8\lambda^2}{1+2\alpha e^{-(\lambda+\eta)\phi_I}}
\ee
Notice that the kinetic coupling constant $f_k$ and $V_0$ appear only in the combination (reestablishing $\kappa$) $\alpha=\kappa^2V_0 f_k=V_0 f_k/M_p^2$. Taking into account the dimensionality of the kinetic coupling one can set $f_k=1/M^2$ and then, $\alpha=V_0/(M^2M_p^2)$.
By replacing $\phi_I$ from (\ref{fieldi}) into (\ref{ns1}) and (\ref{r1}) we find the exact analytical expressions for the scalar spectral index and the tensor-to-scalar ratio in terms of the model parameters and the number of $e$-foldings in the slow-roll approximation:
\be\label{nsexplicit}
n_s=1-\frac{\lambda^2+\lambda(2\lambda+\eta)W\left[\frac{1}{2}\left(\lambda^2-2\right)e^{\frac{\lambda^2}{2}+\lambda N(\lambda+\eta)-1}\right]}{\Big(1+W\left[\frac{1}{2}\left(\lambda^2-2\right)e^{\frac{\lambda^2}{2}+\lambda N(\lambda+\eta)-1}\right]\Big)^2}
\ee
\be\label{rexplicit}
r=\frac{8\lambda^2}{1+W\left[\frac{1}{2}\left(\lambda^2-2\right)e^{\frac{\lambda^2}{2}+\lambda N(\lambda+\eta)-1}\right]}
\ee
And for the slow-roll parameters $N$ $e$-folds before the end of inflation, we find the following analytical expressions
\be\label{epsilk0}
\epsilon_0=\frac{\lambda^2}{2\Big(1+W\left[\frac{1}{2}\left(\lambda^2-2\right)e^{\frac{\lambda^2}{2}+\lambda N(\lambda+\eta)-1}\right]\Big)}
\ee
\be\label{epsilk1}
\epsilon_1=\frac{\lambda(\lambda+\eta)W\left[\frac{1}{2}\left(\lambda^2-2\right)e^{\frac{\lambda^2}{2}+\lambda N(\lambda+\eta)-1}\right]}{\Big(1+W\left[\frac{1}{2}\left(\lambda^2-2\right)e^{\frac{\lambda^2}{2}+\lambda N(\lambda+\eta)-1}\right]\Big)^2}
\ee
\be\label{deltak0}
\Delta_0=\frac{\lambda^2W\left[\frac{1}{2}\left(\lambda^2-2\right)e^{\frac{\lambda^2}{2}+\lambda N(\lambda+\eta)-1}\right]}{2\Big(1+W\left[\frac{1}{2}\left(\lambda^2-2\right)e^{\frac{\lambda^2}{2}+\lambda N(\lambda+\eta)-1}\right]\Big)^2}
\ee
\be\label{deltak1}
\Delta_1=\frac{\lambda(\lambda+\eta)\Big(W\left[\frac{1}{2}\left(\lambda^2-2\right)e^{\frac{\lambda^2}{2}+\lambda N(\lambda+\eta)-1}\right]-1\Big)}{\Big(1+W\left[\frac{1}{2}\left(\lambda^2-2\right)e^{\frac{\lambda^2}{2}+\lambda N(\lambda+\eta)-1}\right]\Big)^2}
\ee
An interesting result from these equations is that the two observables $n_s$ and $r$ and the slow-roll parameters (at the horizon crossing) do not depend on $\alpha$. So, the behavior of $n_s$ and $r$ is controlled exclusively by the dimensionless constants $\lambda$ and $\eta$ and by the number of $e$-foldings $N$. However $\alpha$ is important to define the scalar field at the beginning and the end of inflation as follows from (\ref{fielde}) and (\ref{fieldi}). Having fixed $\alpha=V_0 f_k/M_p^2=V_0/(M^2M_p^2)$ by the initial conditions on the scalar field, we still have freedom to fix $V_0$ by using the COBE-WMAP normalization \cite{cobe,wmap}, which sets the scale of $M$. 
The restrictions imposed by the COBE-WMAP normalization and the tensor-to-scalar ratio allows to set the set the scales of Hubble parameter and the energy involved in the inflation. From  (\ref{slrt13}) 
\be\label{powersd}
P_{\xi}=A_S\frac{H^2}{(2\pi)^2}\frac{{\cal G}_S^{1/2}}{{\cal F}_S^{3/2}}\sim \frac{H^2}{2(2\pi)^2}\frac{1}{{\cal F}_S}\sim  \frac{H^2}{8\pi^2}\frac{1}{\epsilon_0}
\ee
where we used the limit $(\epsilon_0,\epsilon_1,...)\rightarrow 0$ that gives $A_S\rightarrow 1/2$ and $c_S^2\rightarrow 1$. Taking for instance the case $N=60$, $\lambda=2$, $\eta=1.5$ we find $\epsilon_0\sim 0.0048$ and $r\sim 0.077$. Taking into account the COBE-WMAP normalization  we find
\be\label{cobeh}
P_{\xi}\simeq 2.5\times 10^{-9}\sim \frac{H^2}{8\pi^2}\frac{1}{0.0048}\;\; \Rightarrow H\sim 3\times 10^{-5}M_p\sim 7\times 10^{13} Gev.
\ee
And using the tensor-to-scalar ratio under the same approximations done for $P_S$
\be\label{energyscale}
P_T=rP_S\sim 2\frac{H^2}{\pi^2M_p^2}\sim \frac{2V}{3\pi^2M_p^4}\sim (r) 2.5\times 10^{-9}\sim \;\;\;  \Rightarrow V^{1/4}\sim 7\times 10^{-3}M_p\sim 10^{16}Gev.
\ee
Given $V\sim 3\times 10^{-9}M_p^4$ and $\alpha=10^3$, the mass $M$ takes the value $M\sim 10^{-6}M_p$.
In Fig1 we show the behavior of $n_s$ and $r$ assuming $N=60$ for some numerical values of the constants.
\begin{figure}
\centering
\includegraphics[scale=0.7]{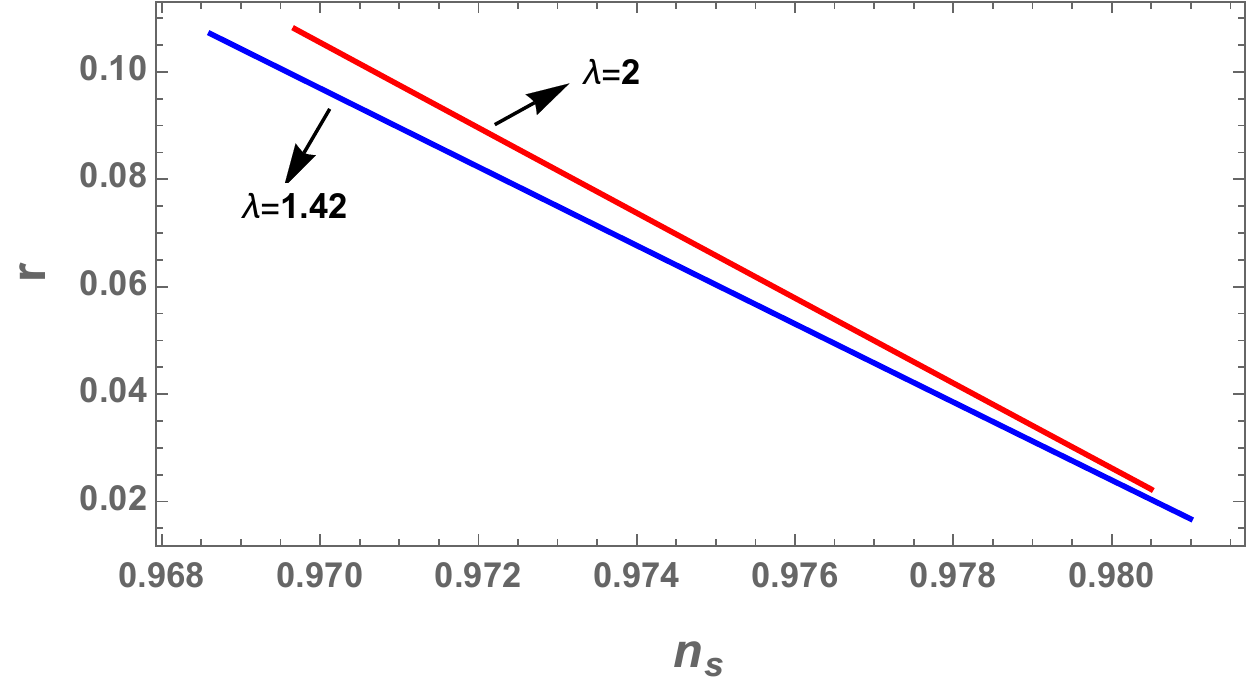}
\caption{$n_s$ vs $r$ for $N=60$, and $\eta$ varying in the interval $1/2<\eta<10$. The red line corresponds to $\lambda=2$ and the blue line to $\lambda=1.42$.}
\end{figure}
In Fig 2 we illustrate the behavior of the slow-roll parameters that show the successful exit from inflation.\\
\begin{figure}
\centering
\includegraphics[scale=0.7]{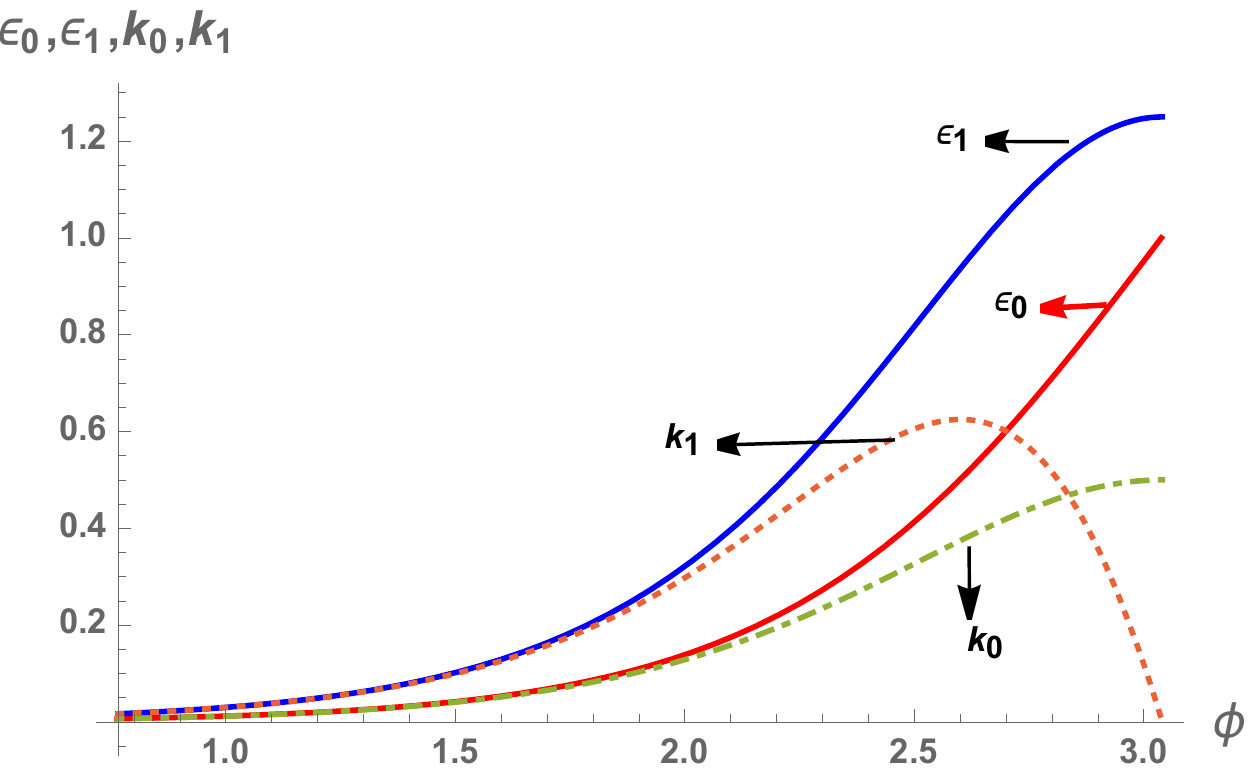}
\caption{The evolution of slow-roll parameters in the interval $\phi_I<\phi<\phi_E$, for $N=60$, $\alpha=10^3$, $\lambda=2$ and $\eta=1/2$.  This behavior allows the exit form inflation. The values of the slow-roll parameters 60 $e$-folds before the end on inflation are:  $\epsilon_0\simeq 0.0067$, $\epsilon_1\simeq 0.016$, $k_0\simeq 0.0067$, $k_1\simeq 0.016$.}
\end{figure}
\noindent A special case takes place when $\eta=-\lambda$. As seen from (\ref{epsilons}) the slow-roll parameters become constant and the model leads to eternal inflation. In this case $N$ and $\phi_I$ are not well defined as follows from (\ref{efolds}) and (\ref{fieldi}) but $n_s$ and $r$ can be found from (\ref{ns1}), (\ref{r1}) and become constants given by
\be\label{nsreternal}
n_s=1-\frac{\lambda^2}{2\alpha+1},\;\;\; r=\frac{8\lambda^2}{2\alpha+1}
\ee
which gives the relationship
\be\label{nsvsr}
n_s=1-\frac{1}{8}r,
\ee
which imply that $n_s$ and $r$ can not simultaneously satisfy the observational restrictions, and therefore this case is discarded. \\

\noindent {\bf Gauss-Bonnet coupling.}\\
\be\label{expogb}
F\left( \phi  \right) = \frac{1}{{{\kappa ^2}}},\,\,\,\,\,\,\,\,\,V\left( \phi  \right) = V_0 e^{-\lambda\kappa\phi},\,\,\,\,\,\,\,\,\,{F_1}\left( \phi  \right) =0 ,\,\,\,\,\,\,\,\,{F_2}\left( \phi  \right) = f_{g} e^{-\eta\kappa\phi}.
\ee
which from (\ref{eqm8}) gives the following slow-roll parameters
\be\nonumber
\epsilon_0=\frac{1}{6}\lambda\left(8\eta\beta e^{-(\lambda+\eta)\phi}+3\lambda\right),\;\;\; \epsilon_1=-\frac{8}{3}\beta\eta\left(\lambda+\eta\right)e^{-(\lambda+\eta)\phi}
\ee
\be\label{epsilong}
\Delta_0=-\frac{8}{9}\beta\eta\left(3\lambda e^{(\lambda+\eta)\phi}+8\beta\eta\right)e^{-2(\lambda+\eta)\phi},\;\;\; \Delta_1=-\frac{1}{3}\left(\lambda+\eta\right)\left(3\lambda e^{(\lambda+\eta)\phi}+16\beta\eta\right)e^{-(\lambda+\eta)\phi}.
\ee
where $\kappa=1$ and $\beta=V_0f_g$. The scalar field at the end of inflation, from the condition $\epsilon_0=1$, takes the form
\be\label{fieldeg}
\phi_E=\frac{1}{\lambda+\eta}\ln\left[\frac{8\beta\eta\lambda}{3(2-\lambda^2)}\right]
\ee
And the number of $e$-foldings from (\ref{eqm17}) is given by
\be\label{efoldsg}
N=\frac{1}{\lambda(\lambda+\eta)}\left(\ln\left[\frac{16\beta\eta}{2-\lambda^2}\right]-\ln\left[8\beta\eta+3\lambda e^{(\lambda+\eta)\phi}\right]\right).
\ee
Solving this equation with respect to the scalar field we find the scalar field $N$ $e$-folds before the end of inflation as
\be\label{fieldig}
\phi_I=\frac{1}{\lambda+\eta}\ln\left[\frac{16\beta\eta e^{-\lambda N(\lambda+\eta)}}{3\lambda(2-\lambda^2)}-\frac{8\beta\eta}{3\lambda}\right]
\ee
Using (\ref{epsilong})  in (\ref{slr29}) gives the expression for scalar spectral index as follows
\be\label{nsg}
n_s=1-\lambda^2+\frac{8}{3}\beta\eta\left(\lambda+2\eta\right)e^{-(\eta+\lambda)\phi_I}.
\ee
And replacing  (\ref{epsilong}) into (\ref{slrt16}) gives the tensor-to-scalar ratio as
\be\label{rg}
r=\frac{8}{9}\left(8\beta\eta+3\lambda e^{(\eta+\lambda)\phi_I}\right)^2 e^{-2(\eta+\lambda)\phi_I}.
\ee
taking into account the above expression for $\phi_I$, it is found
\be\label{nsgi}
n_s=1-\lambda^2-\frac{\lambda(2\eta+\lambda)}{1-\frac{2e^{-\lambda N(\lambda+\eta)}}{2-\lambda^2}},
\ee
and 
\be\label{rgi}
r=\frac{32\lambda^2}{\left(2+(\lambda^2-2)e^{\lambda N(\lambda+\eta)}\right)^2}
\ee
Notice that neither $n_s$ nor $r$ depend on $\beta$, which appears only in the expressions for $\phi_I$ and $\phi_E$. In order to appreciate the order of the parameters involved in inflation, having in mind that at the end of inflation the slow-roll parameters should be of order $1$, we can evaluate the slow-roll parameters at the end of inflation by replacing $\phi_E$ into Eqs. (\ref{epsilong}), giving
\be\label{epsilonge}
\epsilon_0=1,\;\;\; \epsilon_1=\frac{(\lambda+\eta)(\lambda^2-2)}{\lambda},\;\;\; \Delta_0=2-\frac{4}{\lambda^2}, \;\;\; \Delta_1=\frac{(\lambda+\eta)(\lambda^2-4)}{\lambda}.
\ee
Replacing $\Phi_I$ into (\ref{epsilong}) gives the slow-roll parameters $N$ $e$-foldings before the end of inflation as

\be\nonumber
\epsilon_0=\frac{\lambda^2}{2+(\lambda^2-2)e^{N\lambda(\lambda+\eta)}},\;\;\;  \epsilon_1=\frac{\lambda(\lambda+\eta)}{1+\frac{2e^{-N\lambda(\lambda+\eta)}}{\lambda^2-1}}
\ee
\be\label{epsilongi}
\Delta_0=\frac{2\lambda^2(\lambda^2-2)e^{N\lambda(\lambda+\eta)}}{\left(2+(\lambda^2-2)e^{N\lambda(\lambda+\eta)}\right)^2},\;\; \Delta_1=\frac{\lambda(\lambda+\eta)\left((\lambda^2-2)e^{N\lambda(\lambda+\eta)}-2\right)}{2+(\lambda^2-2)e^{N\lambda(\lambda+\eta)}}
\ee

\noindent According to (\ref{epsilonge}), in order to keep $\Delta_0\sim 1$, $\lambda$ should be close to $2$, and from the expressions for $\epsilon_1, \Delta_1$ follows that $\eta\sim -1$. All these approximations are valid under the condition that $\epsilon=1$ at the end of inflation. On the other hand, the exponential in the expressions for $n_s$ and $r$ makes a big difference between $n_s$ and $r$ provided $N\sim 60$ for the above approximations for $\eta$ and $\lambda$. In fact it provides a wrong value for $n_s$ and $r\sim 0$. 
One can also consider the region of parameters where the exponent $e^{-\lambda N(\lambda+\eta)}$ is of order $1$. In this case, numerical analysis shows that if one assumes, for instance the values $\lambda=-0.001$ and $\eta=1$, then $n_s$ and $r$ fall in the appropriate region according to the latest observational data. For $N$ varying in the interval $[50,60]$, $n_s$ and $r$ take values $0.961\le n_s\le 0.967$ and $0.002\le r\le 0.003$.
But in this same interval, the final field (\ref{fieldeg}) which depends on $\eta, \; \lambda,\; \beta$ takes the value $\phi_E\simeq 0.065 M_p$ (assuming $\lambda=-0.001$, $\eta=1$, $\beta=-8\times 10^2$). And the initial field (\ref{fieldig}), which depends  additionally on $N$, varies in the interval $11.6M_p\le \phi_I \le 11.8 M_p$. As we can see the difference between the initial and final fields is almost two orders of magnitude. Besides this, according to (\ref{epsilonge}) when the scalar field reaches the final value, the slow-roll parameters $\epsilon_1, |\Delta_0|, \Delta_1>>1$ ($\epsilon_0=1$) indicating that the slow-roll regime is broken long before the field reaches the value $\phi_E\simeq 0.065 M_p$. 
Numerical analysis shows that at $\phi\simeq 7.6 M_p$ the slow-roll parameters  $\epsilon_1, |\Delta_0|, \Delta_1\sim1$ while $\epsilon<<1$. But given these values of the parameters, for the scalar field it takes $N\approx 1$ to evolve from $\phi_I=11.8 M_p$ to $7.6 M_p$, making the slow-roll mechanism impracticable under the condition $\epsilon_0=1$ at the end of inflation. 
It is also possible to assume that the inflation ends when any of the main slow-roll parameters becomes of order $1$, which in our case would be $\Delta_0$, and have viable inflation (see \cite{sergeioik}). If the condition to end the inflation is imposed on the GB slow-roll parameter $\Delta_0$ (notice that $\epsilon_0$ and $\Delta_0$ enter with the same hierarchy in the expression for the potential (\ref{eqm12})), then the following results can be obtained. First, from the condition $\Delta_0=-1$ the scalar field at the end of inflation takes the value
\be\label{fieldgde}
\phi_E=\frac{1}{\lambda+\eta}\ln\left[\frac{4}{3}\left(\eta\lambda \beta+\sqrt{\beta^2\eta^2(\lambda^2+4)}\right)\right].
\ee
From (\ref{eqm17}) we find the scalar field $N$ $e$-folds before the end of inflation as
\be\label{fieldgid}
\phi_I=\frac{1}{\lambda+\eta}\ln\left[\frac{1}{3\lambda}\left(4e^{-\lambda N(\lambda+\eta)}-8\beta\eta\right)\left(\eta\beta(\lambda^2+2)+\lambda\sqrt{\beta^2\eta^2(\lambda^2+4)}\right)\right].
\ee
This expression for $\phi_I$ leads to the following $n_s$ and $r$ according to (\ref{nsg}) and (\ref{rg}) respectively
\be\label{nsgd}
n_s=1-\lambda^2+\frac{2\beta\eta\lambda(2\eta+\lambda) e^{\lambda N(\lambda+\eta)}}{\lambda\sqrt{\beta^2\eta^2(\lambda^2+4)}+\beta\eta(2-2e^{\lambda N(\lambda+\eta)}+\lambda^2)},
\ee
\be\label{rgd}
r=\frac{8\lambda^2\left(\beta\eta(\lambda^2+2)+\lambda\sqrt{\beta^2\eta^2(\lambda^2+4)}\right)^2}{\left(\lambda\sqrt{\beta^2\eta^2(\lambda^2+4)}+\beta\eta(2-2e^{\lambda N(\lambda+\eta)}+\lambda^2)\right)^2}
\ee
The slow-roll parameters can be also explicitly written in terms of the model parameters when evaluated at the end and at the beginning of inflation.
By replacing (\ref{fieldgde}) into (\ref{epsilong}) we find the following expressions at the end of inflation
\be\nonumber
\epsilon_0=\frac{\lambda\left(\beta\eta(\lambda^2+2)+\lambda\sqrt{\beta^2\eta^2(\lambda^2+4)}\right)}{2\left(\left(\beta\eta\lambda+\sqrt{\beta^2\eta^2(\lambda^2+4)}\right)\right)},\;\; \epsilon_1=-\frac{2\beta\eta(\eta+\lambda)}{\beta\eta\lambda+\sqrt{\beta^2\eta^2(\lambda^2+4)}}
\ee
\be\label{epsilongde}
\Delta_0=-1, \;\; \Delta_1=-\frac{(\eta+\lambda)\left(\beta\eta(\lambda^2+2)+\lambda\sqrt{\beta^2\eta^2(\lambda^2+4)}\right)}{\beta\eta\lambda+\sqrt{\beta^2\eta^2(\lambda^2+4)}}
\ee
And replacing (\ref{fieldgid}) into  (\ref{epsilong}) gives the slow-roll parameters $N$ $e$-foldings before the end of inflation as
\be\label{epsgd0}
\epsilon_0=\frac{\lambda^2\left(\beta\eta(\lambda^2+2)+\lambda\sqrt{\beta^2\eta^2(\lambda^2+4)}\right)}{2\left(\left(\beta\eta(2-2e^{\lambda N(\lambda+\eta)}+\lambda^2)+\lambda\sqrt{\beta^2\eta^2(\lambda^2+4)}\right)\right)},
\ee
\be\label{epsgd1}
\epsilon_1=-\frac{2\beta\eta\lambda(\eta+\lambda) e^{\lambda N(\lambda+\eta)}}{\lambda\sqrt{\beta^2\eta^2(\lambda^2+4)}+\beta\eta(2-2e^{\lambda N(\lambda+\eta)}+\lambda^2)},
\ee
\be\label{delta0}
\Delta_0=-\frac{2\beta\eta\lambda^2\left(\beta\eta(\lambda^2+2)+\lambda\sqrt{\beta^2\eta^2(\lambda^2+4)}\right)e^{N\Lambda(\lambda+\eta)}}{\left(\beta\eta(2-2e^{\lambda N(\lambda+\eta)}+\lambda^2)+\lambda\sqrt{\beta^2\eta^2(\lambda^2+4)}\right)^2},
\ee
\be\label{delta1}
\Delta_1=-\frac{\lambda(\lambda+\eta)\left(\beta\eta(\lambda^2+2+2e^{N\lambda(\lambda+\eta)})+\lambda\sqrt{\beta^2\eta^2(\lambda^2+4)}\right)}{\lambda\sqrt{\beta^2\eta^2(\lambda^2+4)}+\beta\eta(2-2e^{\lambda N(\lambda+\eta)}+\lambda^2)}.
\ee
Assuming for instance, $N=50$, $\lambda=-0.005$, $\eta=1$, $f_2=-1$, we find
$$n_s=0.965,\;\; r=0.004,\;\; \phi_I=5.06,\;\; \phi_E=0.988.$$
The slow-roll parameters $50$ $e$-foldings before the end of inflation take the values
\be\label{sample}
(\epsilon_0, \; \epsilon_1,\; \Delta_0,\; \Delta_1)\big|_{\phi_I}=(0.000056, 0.017, -0.00039, 0.039),
\ee
and at the end of inflation
\be\label{sample1}
(\epsilon_0, \; \epsilon_1,\; \Delta_0,\; \Delta_1)\big|_{\phi_E}=(0.0025, 0.99, -1, 1.99).
\ee
In Fig.3 we show the $n_s-r$ trajectory for $50\le N\le 60$.
\begin{figure}
\centering
\includegraphics[scale=0.7]{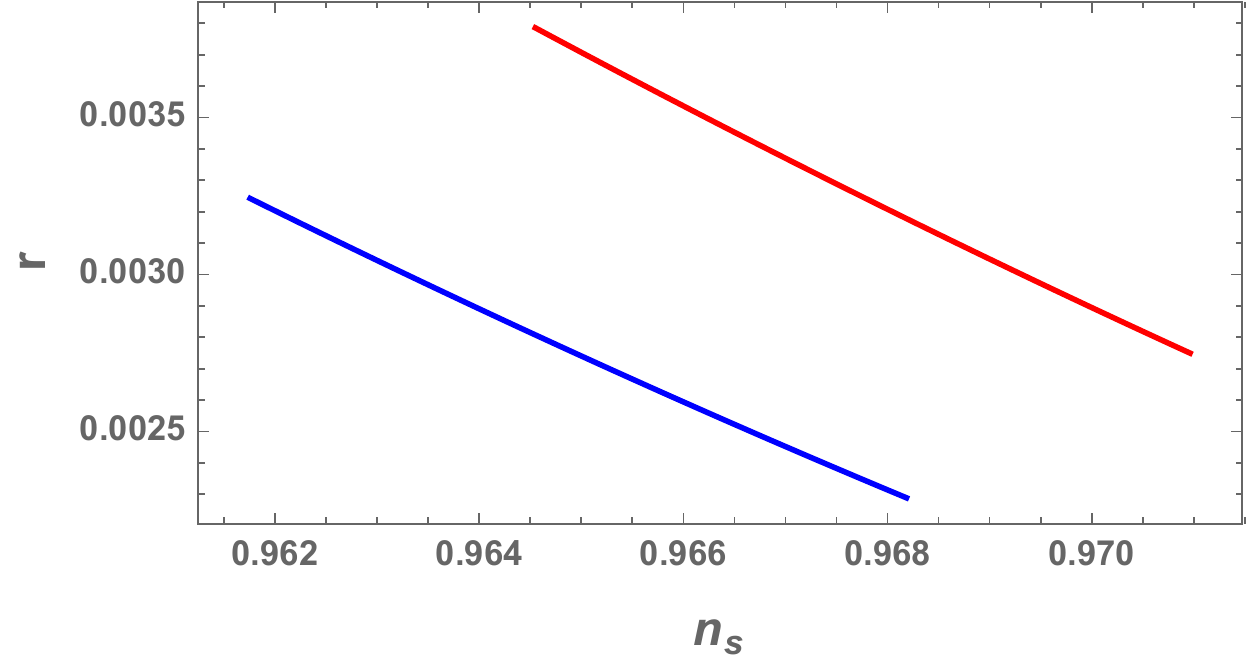}
\caption{The scalar spectral index $n_s$ and tensor/scalar ratio $r$, for  $\lambda=-0.001,\; \eta=1,\; \beta=-1$ (blue) and  $\lambda=-0.004,\; \eta=1,\; \beta=-1$ (red), for $N$ varying between $50\le N\le 60$. Both curves fall in the region constrained by the latest observations.}
\end{figure}
\noindent The above results show that the scalar potential in the frame of scalar-tensor models is not the only magnitude that drives the inflation. The  effect of interactions terms amounts to the effect of an effective potential since the scalar field also rolls down the coupling functions. 
Concerning the restrictions imposed by the COBE-WMAP normalization, we find from  (\ref{slrt13}) 
\be\label{powersd}
P_{\xi}=A_S\frac{H^2}{(2\pi)^2}\frac{{\cal G}_S^{1/2}}{{\cal F}_S^{3/2}}\sim \frac{H^2}{2(2\pi)^2}\frac{1}{{\cal F}_S}\sim  \frac{H^2}{(2\pi)^2}\frac{1}{2\epsilon_0-\Delta_0}
\ee
Taking into account the values for the sample (\ref{sample}), where $\Delta_0$ is larger than $\epsilon$, and using the COBE normalization for the power spectrum $P_{\xi}$, we can write
\be\label{hubblescale}
P_{\xi}\simeq 2.5\times 10^{-9}\sim \frac{H^2}{(2\pi)^2}\frac{1}{4\times 10^{-4}}\;\;\;  \Rightarrow H\sim 6.3\times 10^{-6}M_p\sim 10^{13}Gev. 
\ee
And from the tensor-to-scalar ratio it is found
\be\label{energyscale}
P_T=rP_S\sim 2\frac{H^2}{\pi^2M_p^2}\sim \frac{2V}{3\pi^2M_p^4}\sim (r) 2.5\times 10^{-9} \;\;\;  \Rightarrow V^{1/4}\sim 3\times 10^{-3}M_p\sim 7\times 10^{15}Gev,
\ee
where we used the value for $r$ given in the sample (\ref{sample1}). This imply, given that $\beta=\frac{f_gV_0}{M_p^4}=-1$, that the GB coupling constant $|f|\sim 10^{12}$. \\
A special case takes place when $\eta=-\lambda$ in (\ref{expogb}). The slow-roll parameters become constants given by
\be\label{epsilconst}
\epsilon_0=\frac{1}{6}\left(3-8\beta\right)\lambda^2,\;\; \epsilon_1=0,\;\; \Delta_0=\frac{8}{9}\left(3-8\beta\right)\beta\lambda^2
\ee
The spectral index and tensor-to-scalar ratio are given by
\be\label{nsr}
n_s=1-\left(1-\frac{8}{3}\beta\right)\lambda^2,\;\;\; r=8\left(1-\frac{8}{3}\beta\right)^2\lambda^2
\ee
Solving these equations with respect to $\beta$ and $\lambda$ gives
\be\label{betalambda}
\beta=\frac{3(8n_s+r-8)}{64(n_s-1)},\;\;\; \lambda=\pm\frac{2\sqrt{2}(1-n_s)}{\sqrt{r}}.
\ee
Thus, given the values of the observables $n_s$ and $r$, we can find the model parameters. Taking for instance $n_s=0.968$ and $r=10^{-2}$, give $\beta\simeq 0.36$ and $\lambda\simeq \pm 0.9$. According to (\ref{hubblescale}) and (\ref{energyscale})
$$P_{\xi}\sim  \frac{H^2}{(2\pi)^2}\frac{1}{2\epsilon_0-\Delta_0}\sim  \frac{H^2}{(2\pi)^2}\frac{1}{1.3\times 10^{-3}}\sim \simeq 2.5\times 10^{-9}\Rightarrow H\sim 10^{-5}M_p$$
and 
$$P_T=rP_S\sim \frac{2V}{3\pi^2M_p^4}\sim (10^{-2}) 2.5\times 10^{-9}\;\; \Rightarrow V^{1/4}\sim 4.4\times 10^{-3}M_p$$\\
The issue with this case is the constancy of the slow-roll parameters that leads to eternal inflation, unless an alternative mechanism to trigger the graceful exit from inflation is provided.\\

\noindent {\bf Kinetic and Gauss-Bonnet couplings I.}\\

\noindent The following model includes both, the non-minimal kinetic and Gauss-Bonnet couplings.\\
\be\label{expokgb}
F\left( \phi  \right) = \frac{1}{{{\kappa ^2}}},\,\,\,\,\,\,\,\,\,V\left( \phi  \right) = V_0 e^{-\lambda\kappa\phi},\,\,\,\,\,\,\,\,\,{F_1}\left( \phi  \right) =f_k e^{\lambda\kappa\phi},\,\,\,\,\,\,\,\,{F_2}\left( \phi  \right) = f_{g} e^{-\lambda\kappa\phi}.
\ee
The slow-roll parameters in terms of the scalar field take the form
\be\nonumber
\epsilon_0=\frac{(8\beta e^{-2\lambda\phi}+3)\lambda^2}{12\alpha+6},\;\;\; \epsilon_1=-\frac{16\beta\lambda^2 e^{-2\lambda\phi}}{6\alpha+3}
\ee
\be\nonumber
\Delta_0=-\frac{8\beta e^{-4\lambda\phi}(8\beta +3e^{2\lambda\phi})\lambda^2}{9(2\alpha+1)},\;\;\; \Delta_1=-\frac{2(16\beta e^{-2\lambda\phi}+3)\lambda^2}{6\alpha+3},
\ee
\be\label{epsilonkgb}
k_0=\frac{\beta e^{-4\lambda\phi}(8\beta +3e^{2\lambda\phi})^2\lambda^2}{9(2\alpha+1)^2},\;\;\; k_1=-\frac{32\beta\lambda^2 e^{-2\lambda\phi}}{6\alpha+3}
\ee
Where $\alpha$ and $\beta$ are defined as before, i.e.  $\alpha=f_k V_0$ and $\beta=f_g V_0$. By solving the condition to end the inflation, $\epsilon_0=1$ we find
\be\label{phiendkgb}
\phi_E=\frac{1}{2\lambda}\ln\left[\frac{8\beta\lambda^2}{3(4\alpha-\lambda^2+2)}\right].
\ee
From (\ref{eqm17}) we find
\be\label{efoldskgb}
N=\frac{3}{\lambda^2(3-8\beta)}\left(\lambda\phi-\alpha e^{-2\lambda\phi}\right)\Big|_{\phi_I}^{\phi_E},
\ee
which gives the scalar field $N$ $e$-folds before the end of inflation as
\be\label{phiinikgb}
\begin{aligned}
\phi_I=&\frac{1}{2\lambda}\left(\ln\left[\frac{8\beta\lambda^2}{3(4\alpha-\lambda^2+2)}\right]+W\left[\frac{3\alpha(4\alpha-\lambda^2+2)}{4\beta\lambda^2} e^{\frac{36\alpha^2+8N\beta(3-8\beta)\lambda^4-9\alpha(\lambda^2-2)}{12\beta\lambda^2}}\right]\right)\\ &
+\frac{9\alpha\left(\lambda^2-4\alpha-2\right)+8N\beta\lambda^4(8\beta-3)}{24\beta\lambda^3}
\end{aligned}
\ee
writing the scalar spectral index in terms of the scalar field from (\ref{slr29}) and using  (\ref{epsilonkgb}) we find 
\be\label{nskgb}
n_s=1+\frac{\left(8\beta e^{-2\lambda\phi}-1\right)\lambda^2}{2\alpha+1}
\ee
and for the tensor-to-scalar ratio from (\ref{slrt16}) and (\ref{epsilonkgb}) we find
\be\label{rkgb}
r=\frac{8\lambda^2\left(3 e^{2\lambda\phi}+8\beta\right)^2 e^{-4\lambda\phi}}{9(2\alpha+1)}.
\ee
At the horizon crossing, $N$ $e$-folds before the end of inflation, we find the following expressions for $n_s$ and $r$
\be\label{nskgbhc}
n_s=1-\frac{\lambda^2}{2\alpha+1}+\frac{4\beta\lambda^2}{\alpha(2\alpha+1)}W\left[\frac{3\alpha(4\alpha-\lambda^2+2)}{4\beta\lambda^2} e^{\frac{36\alpha^2+8N\beta(3-8\beta)\lambda^4-9\alpha(\lambda^2-2)}{12\beta\lambda^2}}\right],
\ee
and
\be\label{rkgbhc}
r=\frac{8\lambda^2}{9\alpha^2(2\alpha+1)}\left(3\alpha+4\beta W\left[\frac{3\alpha(4\alpha-\lambda^2+2)}{4\beta\lambda^2} e^{\frac{36\alpha^2+8N\beta(3-8\beta)\lambda^4-9\alpha(\lambda^2-2)}{12\beta\lambda^2}}\right]\right)^2
\ee
Notice that setting $\beta=0$ we obtain the previous results (\ref{nsreternal}) for $n_s$ and $r$. Evaluating the slow-roll parameters at the end of inflation (under the condition $\epsilon=1)$ we find
\be\nonumber
\epsilon_0=1,\epsilon_1=\frac{6(\lambda^2-4\alpha-2)}{6\alpha+3},\;\; \Delta_0=\frac{2(\lambda^2-4\alpha-2)}{\lambda^2}
\ee
\be\label{epsilonkgbhc}
\Delta_1=\frac{2(\lambda^2-8\alpha-4)}{2\alpha+1},\;\; k_0=\frac{4\alpha}{\lambda^2},\;\; k_1=\frac{4(\lambda^2-4\alpha-2)}{2\alpha+1}
\ee
Looking at the expressions (\ref{nskgbhc}) and (\ref{rkgbhc}) it can be seen that to obtain the observable values for $n_s$ and $r$, $\lambda$ should be small or of the order $1$ and $\alpha$ should be large. But, according to (\ref{epsilonkgbhc}), this will make $\Delta_0>>1$ and $k_0>>1$, meaning that they become of the order $1$ long before the end of inflation, spoiling the slow-roll approximation. Therefore it is not possible to satisfy the condition that all slow-roll parameters maintain in the region of $\pm 1$ at the end of inflation, for appropriate values of $\lambda$ and $\alpha$. Better results are obtained with the following model.\\

\noindent {\bf Kinetic and Gauss-Bonnet couplings II.}\\
\be\label{expokgb1}
F\left( \phi  \right) = \frac{1}{{{\kappa ^2}}},\,\,\,\,\,\,\,\,\,V\left( \phi  \right) = V_0 e^{-\lambda\kappa\phi},\,\,\,\,\,\,\,\,\,{F_1}\left( \phi  \right) =f_k e^{-\lambda\kappa\phi},\,\,\,\,\,\,\,\,{F_2}\left( \phi  \right) = f_{g} e^{\lambda\kappa\phi}.
\ee
where the slow-roll parameters in terms of the scalar field take the form
\be\nonumber
\epsilon_0=\frac{(3-8\beta)\lambda^2 e^{2\lambda\phi}}{6(e^{2\lambda\phi}+2\alpha)},\;\;\; \epsilon_1=\frac{4\alpha(3-8\beta)\lambda^2 e^{2\lambda\phi}}{3(e^{2\lambda\phi}+2\alpha)^2}
\ee
\be\nonumber
\Delta_0=\frac{8\beta(3-8\beta)\lambda^2 e^{2\lambda\phi}}{9(e^{2\lambda\phi}+2\alpha)},\;\;\; \Delta_1=\frac{4\alpha(3-8\beta)\lambda^2 e^{2\lambda\phi}}{3(e^{2\lambda\phi}+2\alpha)^2},
\ee
\be\label{epsilonkgb1}
k_0=\frac{\alpha(3-8\beta)^2\lambda^2 e^{2\lambda\phi}}{9(e^{2\lambda\phi}+2\alpha)^2},\;\;\; k_1=-\frac{2(3-8\beta)\lambda^2 e^{2\lambda\phi}(e^{2\lambda\phi}-2\alpha)}{3(e^{2\lambda\phi}+2\alpha)^2}
\ee
The end of inflation takes place for the scalar field $\phi_E$ given by
\be\label{phiendkgb1}
\phi_E=\frac{1}{2\lambda}\ln\left[\frac{12\alpha}{\lambda^2(3-8\beta)-6}\right]. 
\ee
The number of $e$-folds, from  (\ref{eqm17}) is given by
\be\label{efoldskgb1}
N=\frac{3}{\lambda^2(3-8\beta)}\left(\lambda\phi-\alpha e^{-2\lambda\phi}\right)\Big|_{\phi_I}^{\phi_E}.
\ee
Replacing $\phi_E$ and solving with respect to $\phi_I$ we find
\be\label{phiinikgb1}
\begin{aligned}
\phi_I=&\frac{1}{12\lambda}\Big(6\ln\left[\frac{12\alpha}{\lambda^2(3-8\beta)-6}\right]+6W\left[\frac{1}{6}\left((3-8\beta)\lambda^2-6\right) e^{\frac{1}{6}(3-8\beta)(1+4N)\lambda^2-1}\right]\\ &+\lambda^2\left(32\beta N-12N+8\beta-3\right)+6\Big).
\end{aligned}
\ee
The scalar spectral index in terms of the scalar field is given by the following expression (from (\ref{slr29}) and (\ref{epsilonkgb1}))
\be\label{nsphikgb}
n_s=\frac{12\alpha^2+6\alpha((8\beta-3)\lambda^2+2)e^{2\lambda\phi}+((8\beta-3)\lambda^2+3))e^{4\lambda\phi}}{3(e^{2\lambda\phi}+2\alpha)^2},
\ee
and for the tensor/scalar ratio (using (\ref{slrt16}) and (\ref{epsilonkgb1})) it is found
\be\label{rphikgb}
r=\frac{8\lambda^2(3-8\beta)^2 e^{2\lambda\phi}}{9(e^{2\lambda\phi}+2\alpha)}.
\ee
The observed values of $n_s$ and $r$ are found through the evaluation of the above expressions $N$ $e$-foldings before the end of inflation, leading to 
\be\label{nskgb1}
n_s=1-\frac{(3-8\beta)\lambda^2\Big(1+3W\left[\frac{1}{6}\left((3-8\beta)\lambda^2-6\right) e^{\frac{1}{6}(3-8\beta)(1+4N)\lambda^2-1}\right]\Big)}{3\Big(1+W\left[\frac{1}{6}\left((3-8\beta)\lambda^2-6\right) e^{\frac{1}{6}(3-8\beta)(1+4N)\lambda^2-1}\right]\Big)^2}
\ee
and 
\be\label{rkgb1}
r=\frac{8\lambda^2(3-8\beta)^2}{9\Big(1+W\left[\frac{1}{6}\left((3-8\beta)\lambda^2-6\right) e^{\frac{1}{6}(3-8\beta)(1+4N)\lambda^2-1}\right]\Big)}
\ee
Notice that in fact the dependence of $n_s$ and $r$ on $\alpha$ disappears when evaluated at the horizon crossing. Replacing (\ref{phiendkgb1}) into
(\ref{epsilonkgb1}) we find the expressions for the slow-roll parameters at the end of inflation 
\be\label{epsikgbend} 
\begin{aligned}
\epsilon_0=1,\;\; \epsilon_1=&4+\frac{24}{(8\beta-3)\lambda^2},\;\; \Delta_0=\frac{16\beta}{3},\;\; \Delta_1=4+\frac{24}{(8\beta-3)\lambda^2},\\
&  k_0=1-\frac{2}{\lambda^2}-\frac{8\beta}{3},\;\; k_1=4+\frac{48}{(8\beta-3)\lambda^2}.
\end{aligned}
\ee
Analyzing these results, it can be seen that it is possible to find values $\epsilon_1, \Delta_0, ...\sim 1$ (guaranteeing graceful exit from inflation), assuming $|f_2|<<1$ and $\lambda\sim 1$, which at the same time give adequate values for the observables $n_s$ and $r$. In Fig. 4 we show the evolution of $n_s$ and $r$ for the number of $e$-foldings in the interval $50\le N\le 60$.
\begin{figure}
\centering
\includegraphics[scale=0.7]{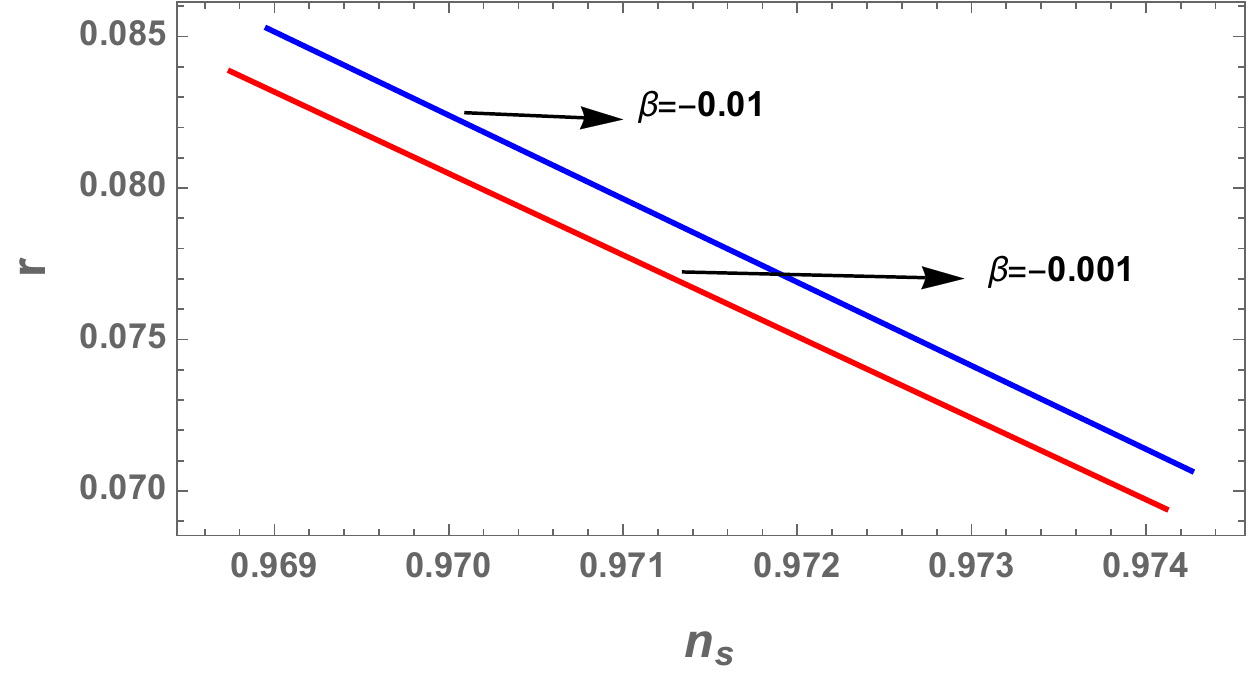}
\caption{The scalar spectral index $n_s$ and tensor/scalar ratio $r$, for  $\lambda=1.42,\; \beta=-0.01$ (blue) and  $\lambda=1.42,\; \beta=-0.001$ (red), for $N$ varying between $50\le N\le 60$. $n_s$ and $r$ do not depend on $\alpha$, which is used to set the values of $\phi_I$ and $\phi_E$.} 
\end{figure}
taking for instance $N=60,\; \lambda=1.42,\; \alpha=10^3,\; \beta=-0.001$, the slow-roll parameters at the beginning of inflation take the values
$$
\epsilon_0\simeq 0.0043,\; \epsilon_1\simeq 0.017,\; \Delta_0\simeq -0.000023,\; \Delta_1\simeq 0.017,\; k_0\simeq 0.0043,\; k_1\simeq 0.017.
$$
\noindent Following the same lines as in the previous cases, we can evaluate the size of the Hubble parameter and the energy involved during inflation, obtaining that $H\sim 3\times 10^{-5} M_p$ and $V^{1/4}\sim 7\times 10^{-3} M_p$ (taking into account the above slow-roll parameters).
The evolution of the slow-roll parameters for this case is shown in Fig. 5, where $\phi_I\simeq 0.76 M_p$ and $\phi_E\simeq 4.2 M_p$.
 \begin{figure}
\centering
\includegraphics[scale=0.7]{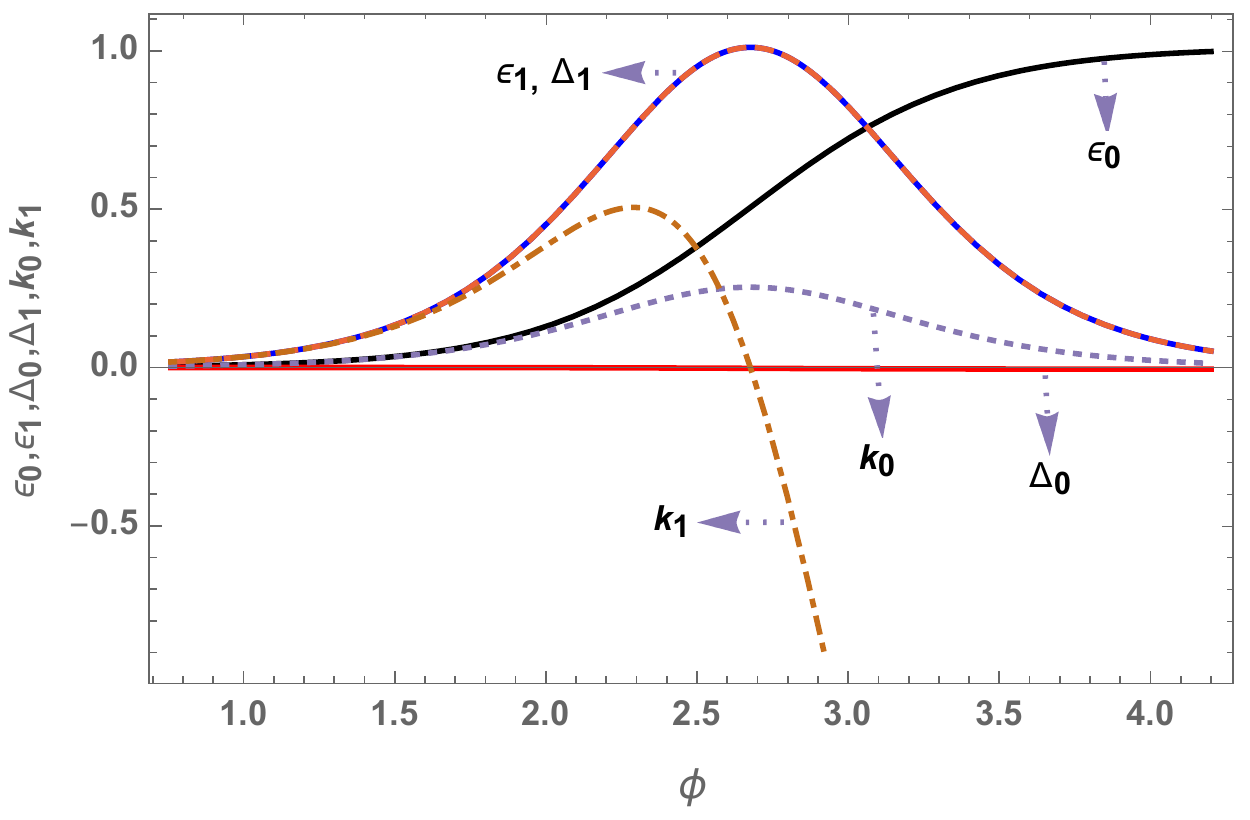}
\caption{The variation of the slow-roll parameters between $\phi_I\simeq 0.76 M_p$ and $\phi_E\simeq 4.2 M_p$ obtained for $N=60,\; \lambda=1.42,\; \alpha=10^3,\; \beta=-0.001$.} 
\end{figure}
Observing Fig. 5 we can see that the slow-roll dynamics can also be consistent if one imposes the condition to end the inflation on the slow-roll parameters $\epsilon_1=\Delta_1=1$. This leads (from (\ref{epsilonkgb1})) to the following scalar field at the end of inflation
\be\label{phiendkgb2}
\phi_E=\frac{1}{2\lambda}\ln\left[\frac{2}{3}\left(\alpha((3-8\beta)\lambda^2-3)+\sqrt{\alpha^2\lambda^2(8\beta-3)(6+(8\beta-3)\lambda^2)}\right)\right]
\ee
Then, from (\ref{efoldskgb1}) and replacing $\phi_E$ given by (\ref{phiendkgb2}) it is found
\be\label{phiikgb2}
\phi_I=\frac{1}{3}(8\beta-3)(N-f_N)\lambda+\frac{1}{2\lambda}W\left[2\alpha e^{-\frac{2}{3}(8\beta-3)(N-f_N)\lambda^2}\right],
\ee
which leads, from (\ref{nsphikgb}) and (\ref{rphikgb}), to the following expressions for $n_s$ and $r$
\be\label{nskbg2}
n_s=1-\frac{(3-8\beta)\lambda^2\Big(1+3W\Big[2\alpha e^{-\frac{2}{3}(8\beta-3)(N-f_N)\lambda^2}\Big]\Big)}{3\Big(1+3W\Big[2\alpha e^{-\frac{2}{3}(8\beta-3)(N-f_N)\lambda^2}\Big]\Big)^2}
\ee
and
\be\label{rkgb2}
r=\frac{8\lambda^2}{9\alpha^2(2\alpha+1)}\left(3\alpha+4\beta W\left[2\alpha e^{-\frac{2}{3}(8\beta-3)(N-f_N)\lambda^2}\right]\right)^2
\ee
Fig. 6 shows the behavior of  $n_s$ and $r$ for $50\le N\le 60$.
\begin{figure}
\centering
\includegraphics[scale=0.7]{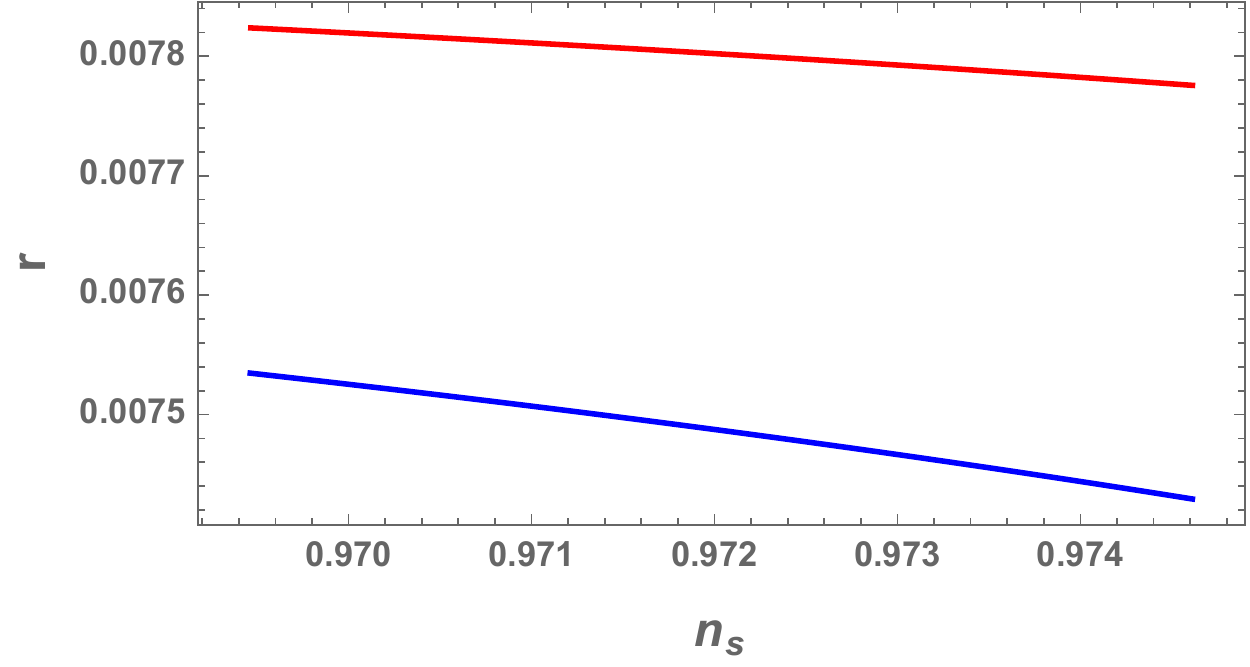}
\caption{The scalar spectral index $n_s$ and tensor/scalar ratio $r$, for  $\lambda=1.42,\; \alpha=10^3,\; \beta=-0.1$ (blue) and  $\lambda=1.42,\; \alpha=10^3, \; \beta=-0.05$ (red), for $N$ in the interval $50\le N\le 60$. The tensor/scalar ratio is an order of magnitude smaller than the case depicted in Fig.4.} 
\end{figure}
The variation of the slow-roll parameters between the beginning and the end of inflation is shown in fig 7.
\begin{figure}
\centering
\includegraphics[scale=0.7]{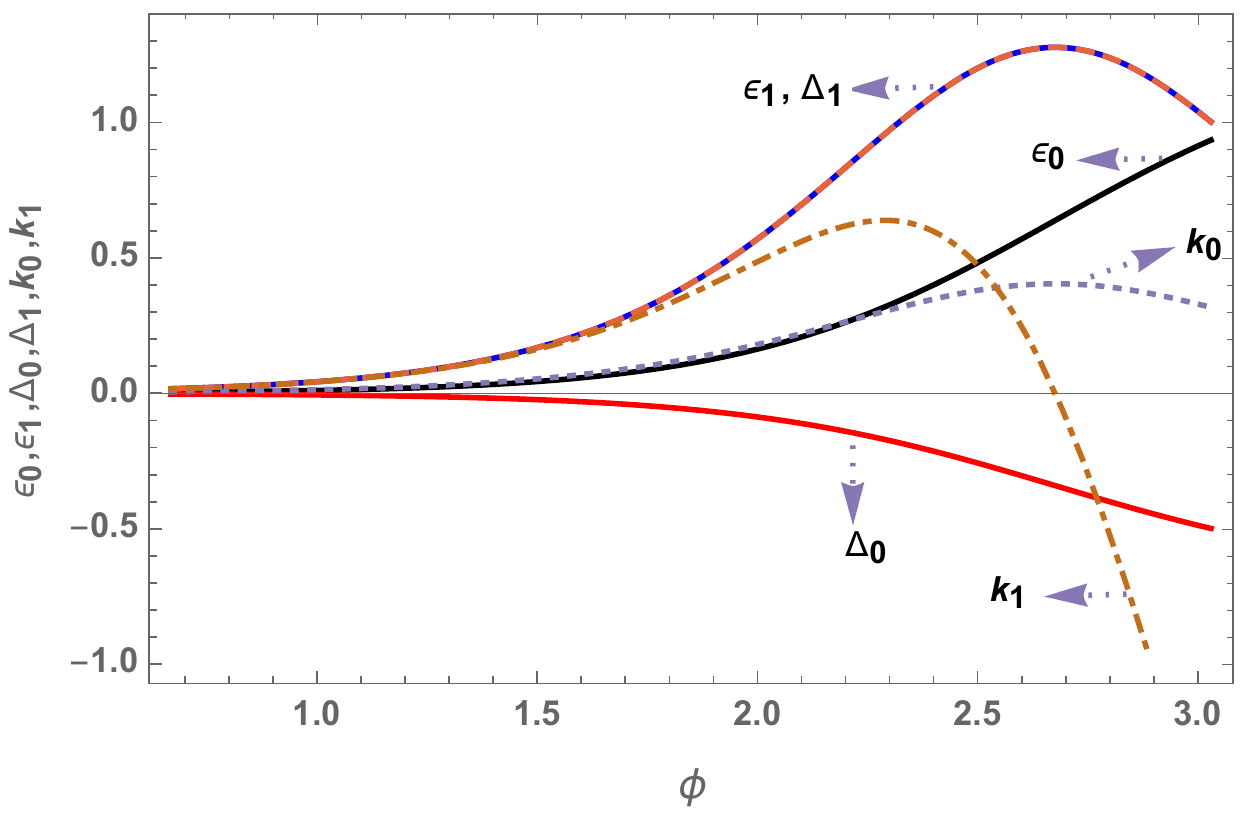}
\caption{The variation of slow-roll parameters between $\phi_I\simeq 0.67 M_p$ and $\phi_E\simeq 3 M_p$ obtained for $60$ $e$-foldings, with $\lambda=1.42,\; \alpha=10^3,\; \beta=-0.1$. The growth towards values of the order of $\pm 1$ is more homogeneous than the one depicted in Fig.5} 
\end{figure}
The values of the slow-roll parameters $60$ $e$-folds before the end of inflation ($\lambda=1.42,\; \alpha=10^3,\; \beta=-0.1$) are
$$
\epsilon_0\simeq 0.0042,\; \epsilon_1\simeq 0.017,\; \Delta_0\simeq -0.0023,\; \Delta_1\simeq 0.017,\; k_0\simeq 0.005,\; k_1\simeq 0.017.
$$
\noindent The Hubble and energy scales involved in the process of inflation, for the present case, are $H\sim 3\times 10^{-5} M_p$ and $V^{1/4}\sim 4\times 10^{-3} M_p$ .\\

\noindent {\bf Kinetic and Gauss-Bonnet couplings III.}\\
\be\label{expokgb1}
F\left( \phi  \right) = \frac{1}{{{\kappa ^2}}},\,\,\,\,\,\,\,\,\,V\left( \phi  \right) = V_0 e^{-\lambda\kappa\phi},\,\,\,\,\,\,\,\,\,{F_1}\left( \phi  \right) =f_k e^{\lambda\kappa\phi},\,\,\,\,\,\,\,\,{F_2}\left( \phi  \right) = f_{g} e^{\lambda\kappa\phi}.
\ee
This model leads to exact power-law inflation with the constant slow-roll parameters given by
\be\nonumber
\epsilon_0=\frac{(3-8\beta)\lambda^2}{6(\alpha+1)},\;\; \epsilon_1=0,\;\; \Delta_0=\frac{8\beta(3-8\beta)\lambda^2}{9(2\alpha+1)}
\ee
\be\label{epsilonpwl}
\Delta_1=0,\;\; k_0=\frac{\alpha(3-8\beta)^2\lambda^2}{9(2\alpha+1)^2},\;\; k_1=0,
\ee
which predict the scalar spectral index and tensor/scalar ratio given by the following expressions
\be\label{nsr-kgb}
n_s=1-\frac{(3-8\beta)\lambda^2}{3(2\alpha+1)},\;\;\; r=\frac{8(3-8\beta)^2\lambda^2}{9(2\alpha+1)}.
\ee
These equations can be solved with respect to $\alpha$ and $\beta$, resulting in
\be\label{couplings}
 \alpha=\frac{16n_s-8n_s^2+\lambda^2 r-8}{16(n_s-1)^2},\;\;\; \beta=\frac{3(8n_s+r-8)}{64(n_s-1)}
 \ee  
Thus, for a given $\lambda$ we can always find adequate values for $\alpha$ and $\beta$ that satisfy the observed values of $n_s$ and $r$. Taking for instance, $\lambda=2$, $ns=0.968$ and $r=0.01$, the reconstructed couplings acquire the values $\alpha=1.94$ and $\beta=0.36$, and the slow-roll parameters take the values $\epsilon_0\simeq 0.016,\; \Delta_0\simeq 0.0031,\; k_0\simeq 0.00005,\; \epsilon_1=\Delta_1=k_1=0 $. Taking $\lambda=0.5$ gives $\alpha\simeq 304,\; \beta\simeq 0.3$ and the slow-roll parameters  $\epsilon_0\simeq 4\times 10^{-5},\; \Delta_0\simeq 6.4\times 10^{-5},\; k_0\simeq 7.8\times 10^{-6},\; \epsilon_1=\Delta_1=k_1=0 $.

\section{Discussion}
We have analyzed the slow-roll dynamics for the scalar-tensor model with non-minimal kinetic and GB couplings, where the potential and the functional form of the couplings are given by exponential functions of the scalar field. These type of couplings appear in a number of compactifications from higher dimensional fundamental theories  such as supergravity and string theory, where the scalar field encodes the size of the extra dimensions.
In the frame of the standard canonical scalar field, the exponential potential leads to important scaling solutions that describe different epochs of cosmological evolution, including solutions with late time accelerated expansion. It also leads to early time inflationary solutions, though it lacks successful exit  from inflation and leads to tensor-to-scalar ratio larger than the current observational limits. 
With the Introduction of additional interactions like the non-minimal kinetic coupling and GB coupling (GB), we address the above shortcomings of the exponential potential and show that the tensor-to-scalar ratio can be lowered to values that are consistent with latest observational constraints \cite{planck15, planck18} and that the model leads to a graceful exit from inflation.\\
First we considered a model with potential $V_0 e^{-\kappa\lambda\phi}$ and kinetic coupling $f_k e^{-\kappa\eta\phi}$ and have found that the observable magnitudes $n_s$and $r$ do not depend on $\alpha=V_0f_k$, and depend only on the number of $e$-foldings and the exponential powers $\lambda$ and $\eta$. The constants $\alpha$ and $\eta$ can be used to set the values of the scalar field at the end and beginning of inflation, obtaining that $\phi_E\lesssim M_p$. A typical behavior of $n_s$ and $r$ in this case is shown in Fig.1. In the particular case $\eta=-\lambda$, the slow-roll parameters become constant, but the obtained relationship between $n_s$ and $r$ (\ref{nsvsr}) makes it imposible to simultaneously satisfy the observational restrictions, making the model non viable for $\eta=-\lambda$.
In the second case we considered the GB coupling given by $F_2=f_g e^{-\kappa\eta\phi}$, and it was found that, similar to the previous case, neither $n_s$ nor $r$ depend on $\beta=V_0 f_g$, but this parameter can be used to set  $\phi_E$ and $\phi_I$. Considering the region of parameters where $e^{-\lambda N(\lambda+\eta)}\sim 1$ it was found that, for $50\le N\le 60$, $n_s$ and $r$ can take values in the intervals $0.961\le n_s\le 0.967$ and $0.002\le r\le 0.003$, and the scalar field at the end of inflation can be as small as $\phi_E\sim 0.07 M_p$. However, in this case some of the slow-roll parameters become larger than $1$ long before $\epsilon_0\sim 1$, breaking the slow-roll conditions. To fix this problem we have chosen to break the slow-roll conditions when $\Delta_0=-1$, which gives excellent results as sown in Fig. 3 and is consistent with the slow-roll formalism according to the values obtained in (\ref{sample}) and (\ref{sample1}). Considering the case $\eta=-\lambda$ it was found that it leads to constant slow-roll inflation, but contrary to the case of kinetic coupling, it is always possible to find adequate values for the scalar spectral index and the tensor-to-scalar ratio.
In the model with non-minimal kinetic coupling $F_1=f_k e^{\kappa\lambda\phi}$ and GB coupling $F_2=f_g e^{-\kappa\lambda\phi}$ it was found that in order to obtain viable values of $n_s$ (\ref{nskgbhc}) and $r$ (\ref{rkgbhc}), the conditions $\lambda\lesssim 1$ and $\alpha>>1$  should be satisfied, but this imply according to (\ref{epsilonkgbhc}), that some slow-roll parameters reach values $\sim 1$ long before the end of inflation, spoiling the slow-roll approximation. Better result is obtained with the model  $F_1=f_k e^{-\kappa\lambda\phi}$ and $F_2=f_g e^{\kappa\lambda\phi}$, where there is appropriate slow-roll approximation for values of $\lambda$ and $\beta$ that lead to $n_s$ and $r$ in the range quoted by observations, as seen in Figs. 4 and 5. 
But more appropriate behavior of the slow-roll parameters was found if the condition to end the inflation is assumed as $\epsilon_1=\Delta_1=1$. In the proposed numerical example, the scalar-to-tensor ratio decreases to values $r\sim 0.008$ as shown in Fig. 6, and the growth of the slow-roll parameters toward values of the order of $\pm 1$ at the end of inflation is more homogeneous than in the previous case, as seen in Fig. 7. Finally, the model with $V=V_0 e^{-\kappa\lambda\phi}$, $F_1=f_k e^{\kappa\lambda\phi}$ and $F_2=f_g e^{\kappa\lambda\phi}$ was analyzed. This model leads to inflation with constant slow-roll parameters and, as follows from (\ref{nsr-kgb}) and (\ref{couplings}), it is always possible to find adequate values of $\lambda$, $\alpha$ and $\beta$ that give the observational values of $n_s$ and $r$.
In all models considered in the present paper it was possible to find exact analytical expressions for the scalar spectral index and the tensor-to-scalar ratio, which facilitated the analysis. In all considered numerical examples the Hubble and energy scales involved in the inflationary process were of the order of $H\sim 10^{-5} M_p$ and $V^{1/4}\sim 10^{-3}$. \\
The slow-roll analysis for the exponential potential, in the frame of the scalar-tensor theories with non-minimal kinetic and GB couplings, allows to find the scalar spectral index and tensor-to-scalar ratio in the range set by the latest observational data, and lead to successful exit from inflation. 
The advance in the future observations will allow to establish more accurate restrictions on the inflationary models with non-minimal couplings of the type considered in the present model and reaffirm or rule out its viability.



\section*{Acknowledgments}
\noindent This work was supported by Universidad del Valle under project CI 71187.
 DFJ acknowledges support from COLCIENCIAS, Colombia.



\begin{thebibliography}{99}
\bibitem{guth} A. H. Guth, Phys. Rev. D\textbf{23}, 347 (1981).
\bibitem{linde} A. D. Linde, Phys. Lett. B\textbf{108}, 389 (1982).
\bibitem{steinhardt} A. Albrecht. P. J. Steinhardt, Phys. Rev Lett. \textbf{48}, 1220 (1982)
\bibitem{planck13} . A. R. Ade et al., Planck Collaboration (Planck 2013 results. XXII. Constraints on inflation), Astron. and Astrophys. \textbf{571} (2014) A22; arXiv:1303.5082 [astro-ph.CO]
\bibitem{planck15} P. A. R. Ade et al., Planck Collaboration (Planck 2015 results. XX. Constraints on inflation), Astron. and Astrophys. \textbf{594}, (2016) A20;  arXiv:1502.02114 [astro-ph.CO]
\bibitem{planck18}  Y. Akrami et al., Planck Collaboration (Planck 2018 results. X. Constraints on inflation), arXiv:1807.06211 [astro-ph.CO]
\bibitem{bkp} P. A. R. Ade et al., A Joint Analysis of BICEP2/Keck Array and Planck Data, Phys. Rev. Lett. \textbf{114}, 101301 (2015); arXiv:1502.00612 [astro-ph.CO]
\bibitem{revlinde} A. D. Linde, Particle physics and inflationary cosmology, (Harwood, Chur, Switzerland, 1990) Contemp. Concepts Phys. 5, 1 (1990) [hep-th/0503203].
\bibitem{liddle} A. R. Liddle, D. H. Lyth, Phys. Rept. \textbf{231}, 1 (1993); arXiv:astro-ph/9303019
\bibitem{riotto} A. Riotto, DFPD-TH/02/22; arXiv:hep-ph/0210162
\bibitem{lyth}   D. H. Lyth, A. Riotto, Phys. Rept. \textbf{314}, 1 (1999); arXiv:hep-ph/9807278
\bibitem{mukhanov} V. Mukhanov, Physical Foundations of Cosmology (Cambridge University Press, 2005).
\bibitem{baumann} D. Baumann, TASI 2009; arXiv:0907.5424 [hep-th]
\bibitem{nojirioo} S. Nojiri, S.D. Odintsov, V.K. Oikonomou, Phys. Rept.  \textbf{692},  1-104 (2017); arXiv:1705.11098 [gr-qc]
\bibitem{starobinsky1} A. A. Starobinsky and and H. J. Schmidt, Class. Quant. Grav. \textbf{4},  695 (1987).
\bibitem{mukhanov1} V. F. Mukhanov and G. V. Chibisov, JETP Lett. 33 (1981) 532 [Pisma Zh. Eksp. Teor. Fiz. 33 (1981) 549].
\bibitem{starobinsky3}  A. A. Starobinsky, JETP Lett. 30 (1979) 682 [Pisma Zh. Eksp. Teor. Fiz. 30 (1979) 719].
\bibitem{hawking} S. W. Hawking, Phys. Lett. B 115, 295 (1982).
\bibitem{starobinsky2} A. A. Starobinsky, Phys. Lett. B 117 (1982) 175.
\bibitem{guth1} A. H. Guth and S. Y. Pi, Phys. Rev. Lett. 49, 1110 (1982).
\bibitem{bardeen0} J. M. Bardeen, Phys. Rev. D \textbf{22}, 1882 (1980).
\bibitem{bardeen} J. M. Bardeen, P. J. Steinhardt and M. S. Turner, Phys. Rev. D \textbf{28}, 679 (1983).
\bibitem{futumase} T. Futamase and K. I. Maeda, Phys. Rev. D 39, 399 (1989)
\bibitem{unruh} R. Fakir and W. G. Unruh, Phys. Rev. D \textbf{41}, 1783 (1990).
\bibitem{barrow1} J. D. Barrow, Phys. Rev. D{\bf 51}, 2729 (1995).
\bibitem{barrow2} J. D. Barrow, P. Parsons, Phys. Rev. D{\bf 55}, 1906 (1997); arXiv:gr-qc/9607072.
\bibitem{bezrukov} F. L. Bezrukov and M. Shaposhnikov, Phys. Lett. B \textbf{659}, 703 (2008).
\bibitem{picon} C. Armendariz-Picon, T. Damour and V.F. Mukhanov, Phys. Lett. B \textbf{458} (1999) 209 [hep-th/9904075].
\bibitem{ford} L.H. Ford, Phys. Rev. D 40 (1989) 967.
\bibitem{koivisto} T.S. Koivisto and D.F. Mota, JCAP 08 (2008) 021 [arXiv:0805.4229].
\bibitem{mukhanov2} A. Golovnev, V. Mukhanov and V. Vanchurin, JCAP 06 (2008) 009; [arXiv:0802.2068]
\bibitem{kawasaki} M. Kawasaki, M. Yamaguchi and T. Yanagida, Phys. Rev. Lett. 85 (2000) 3572 [hep-ph/0004243].
\bibitem{davies} S.C. Davis and M. Postma, JCAP 03 (2008) 015 [arXiv:0801.4696].
\bibitem{kallosh} R. Kallosh and A. Linde, JCAP 11 (2010) 011
[arXiv:1008.3375].
\bibitem{soda1} S. Kawai, M. Sakagami, J. Soda, Phys. Lett. B{\bf 437}, 284 (1998); arXiv:gr-qc/9802033.
\bibitem{soda2} S. Kawai, J. Soda Phys.Lett. B460, 41 (1999); 	arXiv:gr-qc/9903017.
\bibitem{maldacena} S. Kachru et al., JCAP \textbf{0310}, 013 (2003); arXiv:hep-th/0308055.
\bibitem{kallosh1} R. Kallosh, Lect. Notes Phys. 738 (2008) 119;  [hep-th/0702059]
\bibitem{soda3} M. Satoh, S. Kanno, J. Soda, Phys. Rev. D{\bf 77}, 023526 (2008); arXiv:0706.3585 [astro-ph]
\bibitem{baumann2} D. Baumann, L. McAllister, Inflation and String Theory, Cambridge University Press, 2015; arXiv:1404.2601 [hep-th]
\bibitem{silver} E. Silverstein, D. Tong, Phys. Rev. D\textbf{70}, 103505 (2004); arXiv:hep-th/0310221
\bibitem{silver1} M. Alishahiha, E. Silverstein, D. Tong, Phys. Rev. D\textbf{70}, 123505 (2004); arXiv:hep-th/0404084
\bibitem{chen} X. Chen, JHEP 0508, 045 (2005); arXiv:hep-th/0501184
\bibitem{easson} D.A. Easson, S. Mukohyama and B.A. Powell, Phys. Rev. D \textbf{81}, (2010) 023512 [arXiv:0910.1353] [SPIRES].
\bibitem{linde3} R. Kallosh, A. Linde, D. Roest, J. High Energ. Phys. 11, 198 (2013); arXiv:1311.0472 [hep-th]
\bibitem{ferrara} S. Ferrara, R. Kallosh, A. Linde, M. Porrati, Phys. Rev. D\textbf{88}, 085038 (2013); arXiv:1307.7696 [hep-th]
\bibitem{sergei1} S.D. Odintsov, V.K. Oikonomou, JCAP 04, 041 (2017); arXiv:1703.02853 [gr-qc]
\bibitem{dimopoulos} K. Dimopoulos, C. Owen, JCAP 06, 027 (2017); arXiv:1703.00305 [gr-qc]
\bibitem{akrami} Y. Akrami, R. Kallosh, A. Linde, V. Vardanyan, JCAP 1806, 041 (2018); arXiv:1712.09693 [hep-th] 
\bibitem{nicolis} A. Nicolis, R. Rattazzi, E. Trincherini, Phys. Rev. D\textbf{79}, 064036 (2009); arXiv:0811.2197 [hep-th]
\bibitem{deffayet} C. Deffayet, G. Esposito-Farese, A. Vikman, Phys. Rev. D\textbf{79}, 084003 (2009); arXiv:0901.1314 [hep-th]
\bibitem{kamada} K. Kamada, T. Kobayashi, M. Yamaguchi, J. Yokoyama, Phys. Rev. D\textbf{83}, 083515 (2011); arXiv:1012.4238 [astro-ph.CO]
\bibitem{ohashi} J. Ohashi, S. Tsujikawa, JCAP 1210, (2012) 035; arXiv:1207.4879 [gr-qc]
\bibitem{yokoyama} T. Kobayashi, M. Yamaguchi, J. Yokoyama, Phys. Rev. Lett. \textbf{105}, 231302 (2010); arXiv:1008.0603 [hep-th]
\bibitem{mizuno} S. Mizuno, K. Koyama, Phys. Rev. D\textbf{82}, 103518 (2010); arXiv:1009.0677 [hep-th]
\bibitem{burrage} C. Burrage, C. de Rham, D. Seery, A. J. Tolley, JCAP 1101 014 (2011); arXiv:1009.2497 [hep-th]
\bibitem{kobayashi1} T. Kobayashi, M. Yamaguchi, J. Yokoyama, Prog. Theor. Phys. \textbf{126}, 511 (2011); arXiv:1105.5723 [hep-th]
\bibitem{capozziello1} S. Capozziello, G. Lambiase and H.J. Schmidt, Annalen Phys. \textbf{9}, 39 (2000);  [gr-qc/9906051].
\bibitem{germani} C. Germani, A. Kehagias, Phys. Rev. Lett. \textbf{105},  011302 (2010); arXiv:1003.2635 [hep-ph].
\bibitem{granda3} L. N. Granda, JCAP \textbf{04}, 016 (2011); arXiv:1104.2253 [hep-th]
\bibitem{stsujikawa} S. Tsujikawa, Phys. Rev. D\textbf{85}, 083518 (2012); arXiv:1201.5926 [astro-ph.CO]
\bibitem{nanyang} N. Yang, Q. Fei, Q. Gao, Y. Gong, Class. Quantum Grav. {\bf 33}, 205001 (2016); arXiv:1504.05839 [gr-qc]
\bibitem{gansu} G. Tumurtushaa, arXiv:1903.05354 [gr-qc]
\bibitem{Kanti:1998jd} P.~Kanti, J.~Rizos and K.~Tamvakis, Phys.\ Rev.\ D {\bf 59}  083512 (1999);  gr-qc/9806085
\bibitem{soda4} M. Satoh, J. Soda, JCAP {\bf 0809}, 019 (2008); arXiv:0806.4594 [astro-ph]
\bibitem{Guo:2009uk} Z.~K.~Guo and D.~J.~Schwarz, Phys.\ Rev.\ D {\bf 80} 063523 (2009);
 arXiv:0907.0427 [hep-th].
\bibitem{Guo:2010jr} Z.~K.~Guo and D.~J.~Schwarz, Phys.\ Rev.\ D {\bf 81} 123520 (2010);
  arXiv:1001.1897 [hep-th].
\bibitem{Jiang:2013gza} P.~X.~Jiang, J.~W.~Hu and Z.~K.~Guo, Phys.\ Rev.\ D {\bf 88} 123508 (2013);
  arXiv:1310.5579 [hep-th].
\bibitem{Koh:2014bka} S.~Koh, B.~H.~Lee, W.~Lee and G.~Tumurtushaa, Phys.\ Rev.\ D {\bf 90},  063527 (2014);
  arXiv:1404.6096 [gr-qc].
\bibitem{Kanti:2015pda} P.~Kanti, R.~Gannouji and N.~Dadhich, Phys.\ Rev.\ D {\bf 92},  041302 (2015);
  arXiv:1503.01579 [hep-th].
 \bibitem{soda5} G. Hikmawan, J. Soda, A. Suroso, F. P. Zen, Phys. Rev. D {\bf 93}, 068301 (2016); arXiv:1512.00222 [hep-th]
 \bibitem{vandeBruck:2017voa} C.~van de Bruck, K.~Dimopoulos, C.~Longden and C.~Owen, arXiv:1707.06839 [astro-ph.CO].
 \bibitem{quiang} Q. Wu, T. Zhu, A. Wang, Phys. Rev. D {\bf 97}, 103502 (2018); arXiv:1707.08020 [gr-qc]
\bibitem{sergeioik} S.D. Odintsov, V.K. Oikonomou,  Phys. Rev. D {\bf 98}, 044039 (2018) ; arXiv:1808.05045 [gr-qc]
\bibitem{sergeioik1} S.D. Odintsov, V.K. Oikonomou, Nucl. Phys. B {\bf 929}, 79 (2018); arXiv:1801.10529 [gr-qc]
\bibitem{chakraborty} S. Chakraborty, T. Paul, S. SenGupta, Phys. Rev. D {\bf 98}, 083539 (2018); arXiv:1804.03004 [gr-qc]
\bibitem{tanmoy} S. Nojiri, S.D. Odintsov, V.K. Oikonomou, N. Chatzarakis, T. Paul, Eur. Phys. J. C {\bf 79}, 565  (2019); arXiv:1907.00403 [gr-qc]
\bibitem{fradkin} E. S. Fradkin, A. A. Tseytlin, Phys. Lett. B\textbf{158}, 316 (1985).
\bibitem{gross} D. J. Gross, J. H. Sloan, Nucl. Phys. B\textbf{291} (1987), 41.
\bibitem{tseytlin} R. R. Metsaev, A.A. Tseytlin, Nucl. Phys. B\textbf{293}, 385 (1987).
\bibitem{meissner} K. A. Meissner, Phys. Lett. B {\bf 392}, 298 (1997).
\bibitem{cartier} C. Cartier, J. Hwang, E. J. Copeland, Phys.Rev. D {\bf 64},103504 (2001); arXiv:astro-ph/0106197
\bibitem{nojiriodintsov} S. Nojiri, S. D. Odintsov, M. Sasaki, Phys. Rev. D {\bf 71}, 123509 (2005).
\bibitem{neupane} B. M. Leith and I. P. Neupane, J. Cosmol. Astropart. Phys. {\bf 05}, (2007) 019.
\bibitem{koivisto1} T. Koivisto and D. F. Mota, Phys. Lett. B 644, 104 (2007).
\bibitem{koivisto2} T. Koivisto and D. F. Mota, Phys. Rev. D 75, 023518 (2007).
\bibitem{suchkov} S. V. Sushkov, Phys. Rev. D {\bf 80}, 103505 (2009).
\bibitem{lgranda} L. N. Granda, J. Cosmol. Astropart. Phys. {\bf 07}, 006 (2010).
\bibitem{grandaloaiza} L. N. Granda, E. Loaiza, Phys. Rev. D {\bf 94}, 063528 (2016).
\bibitem{granjimsan} L. N. Granda, D. F. Jimenez, C. Sanchez, Int. J. Mod. Phys. D {\bf 22}, 1350055 (2013). 
\bibitem{sergeinojiri} S.Nojiri, S. D. Odintsov, Phys. Rept. {\bf 505}, 59  (2011).
\bibitem{copeland1} E. J. Copeland, A. R. Liddle and D. Wands, Phys. Rev. D {\bf 57}, 4686 (1998).
\bibitem{copeland2} T. Barreiro, E. J. Copeland and N. J. Nunes, Phys. Rev. D {\bf 61}, 127301 (2000).
\bibitem{abbott} L. Abbott and M. B. Wise, Nucl.Phys. B244 (1984), 541.
\bibitem{lucchin} F. Lucchin and S. Matarrese, Phys.Rev. D32 (1985), 1316.
\bibitem{yokoyama1} J. Yokoyama and K.-i. Maeda, Phys.Lett. B207 (1988), 31.
\bibitem{liddle1} A. R. Liddle, Phys.Lett. B220 (1989) 502.
\bibitem{ratra} B. Ratra, Phys.Rev. D45 (1992) 1913?1952.
\bibitem{grajim} L. N. Granda, D. F. jimenez, arXiv:1905.08349 [gr-qc], to appear in JCAP
\bibitem{cobe} G. F. Smoot et al. Astrophys. J. {\bf 396}, L1-L5 (1992).
\bibitem{wmap} D. N. Spergel et al. [WMAP Collaboration], Astrophys. J. Suppl. {\bf 148}, 175 (2003).

\end{thebibliography}
\end{document}